\documentclass[10pt]{article}
\usepackage{etex}
\usepackage{float}
\usepackage[symbol]{footmisc}
\usepackage[superscript,sort]{cite}
\usepackage{array}
\usepackage[pdftex,colorlinks=true,linkcolor=black,citecolor=black,urlcolor=black,bookmarks=true,breaklinks=true]{hyperref}
\usepackage[pdftex]{graphicx}
\usepackage[usenames,pdftex]{color}
\definecolor{Red}{rgb}{1.0,0.0,0.0}
\usepackage{amsmath,amssymb}

\usepackage{amsthm,amscd,amsxtra,amsfonts,amsmath,amssymb,multirow}
\usepackage{wrapfig}
\usepackage[footnotesize]{caption}
\usepackage[tiny,compact]{titlesec}
\usepackage{ctable}
\usepackage{bigstrut}
\usepackage{adjustbox}
\usepackage{helvet}

\usepackage{xcolor,colortbl}
\usepackage{subfigure}
\usepackage{tikz}
\usetikzlibrary{shapes,arrows}
\usepackage[T1]{fontenc}

\titlespacing*{\section}{0pt}{*0}{*0}
\titlespacing*{\subsection}{0pt}{*0}{*0}
\titlespacing*{\subsubsection}{0pt}{*0}{*0}
\titlespacing{\paragraph}{0pt}{*0}{*1}

\definecolor{MyPurple}{rgb}{1,0,1}

\newcommand{\beq}[1]{\begin{equation} \label{#1}}
\newcommand{\eeq}{\end{equation}}
\newcommand{\barray}{\begin{array}{ll}}
\newcommand{\earray}{\end{array}}

\begin{document}
\pagenumbering{roman}

\clearpage \pagebreak \setcounter{page}{1}
\renewcommand{\thepage}{{\arabic{page}}}

\title{Flexibility-Rigidity Index  for Protein-Nucleic Acid Flexibility and Fluctuation Analysis
}

\author{
Kristopher Opron$^1$,
Kelin Xia$^2$,
  Zach Burton$^1$ and
Guo-Wei Wei$^{3}$ \footnote{ On leave from the Department of Mathematics, Michigan State University}~\footnote{ Address correspondences  to Guo-Wei Wei. E-mail:wei@math.msu.edu}\\
$^1$  Department of Biochemistry and Molecular Biology\\
Michigan State University, MI 48824, USA \\
$^2$ Department of Mathematics \\
Michigan State University, MI 48824, USA\\
$^3$ Mathematical Biosciences Institute\\
The Ohio State University,
Columbus, Ohio 43210, USA
}

\date{\today}
\maketitle

\begin{abstract}

Protein-nucleic acid complexes are important for many cellular processes including the most essential function  such as transcription and translation. For many protein-nucleic acid complexes, flexibility of both macromolecules has been shown to be critical for specificity and/or function. Flexibility-rigidity index (FRI) has been proposed as an accurate and efficient approach for protein flexibility analysis. In this work, we introduce FRI for the flexibility analysis of protein-nucleic acid complexes. We demonstrate that a multiscale strategy, which incorporates multiple kernels to capture various length scales in biomolecular collective motions, is able to significantly improve the  state of art in the flexibility analysis of protein-nucleic acid complexes.   We take the advantage of the  high accuracy and  ${\cal O}(N)$ computational complexity of our multiscale FRI method to investigate the flexibility of large ribosomal subunits, which is difficult to analyze by alternative approaches.  An anisotropic FRI approach, which involves localized  Hessian matrices, is utilized to study the  translocation dynamics in an RNA polymerase.

\end{abstract}
\maketitle

\pagebreak

\section{Introduction}\label{sec:Intro}

Proteins and the nucleic acids, which include  deoxyribonucleic acid (DNA) and ribonucleic acid (RNA), are among the most essential biomolecules for all known forms of life. In cells, proteins have  a wide variety of important functions, including  supporting organism structure,  catalyzing  reactions involved in transcription and the cell cycle,  participating in signal transduction, and  working as immune agents. Nucleic acids typically function in association with proteins and play a crucial role in encoding, transmitting and expressing genetic information. Genetic information is stored  through the nucleic acid sequence, i.e., the order of nucleotides within a DNA or RNA molecule and  transmitted via
transcription and translation processes. 
{Protein rigidity, flexibility and electrostatics strongly correlate to protein structure and function\cite{Anfinsen:1973}. }
The impact of {biomolecular} electrostatics to their structure,  function and dynamics has been  a subject of intensive study. However the importance of biomolecular flexibility and rigidity to their structure and function has been  overlooked. In general, protein rigidity is responsible for protein three-dimensional (3D) equilibrium  geometric  shapes  and structural function in forms of tubulin, collagen, elastin, and keratin, while protein flexibility is an important  factor in all other protein functions \cite{Frauenfelder:1991}. DNA flexibility is an important effect in DNA packing. {Although the flexibility  of biomolecules is often associated with their motion and dynamics}, which are their response to the external stimuli and  die out at the absolute zero temperature, { flexibility is an intrinsic property.}

Biomolecular flexibility and rigidity can be measured directly or indirectly by many experimental approaches, such as  X-ray crystallography,  nuclear magnetic resonance (NMR) and single-molecule force experiments \cite{Dudko:2006}.  In single-molecule force experiments, including optical tweezers and  nanopore force spectroscopy, the intrinsic rupture rate can be a direct measure of the flexibility and rigidity. In the X-ray structure,   Debye-Waller factors, also known as   B-factors or temperature factors, are computed as the uncertainty for each atom in the least square fitting of between the X-ray diffraction data and the theoretical model.  Debye-Waller factors are interpreted as atomic mean-square-fluctuations at the given experimental temperature, and are associated with biomolecular  flexibility and rigidity.  NMR is  known for its ability to analyze biomolecular flexibility and rigidity under physiological conditions, and at various timescales.

The availability of experimental data makes the theoretical study of biomolecular flexibility and rigidity an interesting and important topic, in which quantitative models can be calibrated and validated.   Molecular dynamics (MD) \cite{McCammon:1977} can be used to elucidate biomolecular collective motion and fluctuation. MD is a powerful technique for  the understanding of the conformational landscapes of biomolecules.   However,  biomolecular flexibility and rigidity are intrinsic properties that are better measured at the motionless and fluctuation free state. Therefore, MD is not efficient for biomolecular flexibility and rigidity analysis.  Alternative approaches including   normal mode analysis (NMA)  \cite{Go:1983,Tasumi:1982,Brooks:1983,Levitt:1985}, graph theory \cite{Jacobs:2001} and elastic network model (ENM) \cite{Bahar:1997,Bahar:1998,Atilgan:2001,Hinsen:1998,Tama:2001,LiGH:2002} become the main workhorses for  biomolecular flexibility and rigidity analysis during the past two decades. In analogy to the time-dependent and time-independent Schr\"{o}dinger's equations, these approaches are designed as time-independent counterparts of the corresponding MD methods \cite{JKPark:2013}.  Consequently, a diagonalization of the interaction matrix or Hamiltonian of a biomolecule is a required procedure to obtain biomolecular   eigenmodes and associated eigenvalues, which are further organized to predict the biomolecular temperature factors.	The low order eigenmodes computed from diagonalizing the Kirchhoff matrix or the  Hessian matrix can be interpreted as the slow motions of the biomolecule around the equilibrium state and thus shed light on the  long-time behavior of the biomolecular dynamics beyond the reach of MD simulations  \cite{Tasumi:1982,Brooks:1983,Levitt:1985}.  Tirion argued  that the potential in the NMA can be simplified to retain only the harmonic potential for elasticity, which is the dominant term  in the MD Hamiltonian \cite{Tirion:1996}. Network theory \cite{Flory:1976} has had considerable impact in flexibility analysis. The combination of elasticity and coarse-grained network gives rise to  elastic network model (ENM) \cite{Hinsen:1998}. Many other network based approaches,  including    Gaussian network model (GNM)   \cite{Bahar:1997,Bahar:1998}  and anisotropic network model (ANM) \cite{Atilgan:2001},   have been developed for biomolecular flexibility analysis.

 It has been demonstrated by Yang et al. \cite{LWYang:2008} that  GNM is about one order more efficient than most other flexibility approaches. GNM is also typically more accurate than ANM in B-factor prediction   \cite{JKPark:2013, Opron:2014}. Applications have been demonstrated in stability \cite{Livesay:2004} analysis,  docking simulation  \cite{Gerek:2010},  viral capsids \cite{Rader:2005,Tama:2005} and domain motions of  hemoglobin \cite{CXu:2003}, F1 ATPase \cite{WZheng:2003,QCui:2004}, chaperonin GroEL \cite{Keskin:2002,WZheng:2007} and the ribosome \cite{Tama:2003,YWang:2004}. More details can be found in a few recent reviews \cite{JMa:2005,LWYang:2008,Skjaven:2009,QCui:2010}. A common feature of the above mentioned time-independent methods is that they resort to the matrix diagonalization procedure. The computational complexity of  the matrix diagonalization is typically of the order of ${\cal O}(N^3)$, where $N$ is the number of elements in the matrix. Such a computational complexity calls for new efficient strategies for the flexibility analysis of large biomolecules.

It is well known that NMA and GNM offer poor flexibility analysis for many macromolecules \cite{Kundu:2002,Kondrashov:2007,Hinsen:2008,GSong:2007}.  Park et al. had studied the performance of NMA and GNM methods for three sets of structures \cite{JKPark:2013}.  They found that   both methods fail to work and deliver negative correlation coefficients (CCs) for many structures  \cite{JKPark:2013}.  They have shown that mean correlation coefficients (MCCs) for the NMA  B-factor prediction of small-sized, medium-sized and large-sized  sets of structures  are about  0.480, 0.482 and 0.494, respectively \cite{JKPark:2013,Opron:2014}. The GNM is considerably more accurate and delivers  MCCs of 0.541, 0.550 and 0.529 for the above test sets  \cite{JKPark:2013,Opron:2014}. Indeed, various improvements, including  crystal environment, solvent type and co-factors,   are proposed \cite{Kundu:2002,Kondrashov:2007,Hinsen:2008,GSong:2007}. Additionally,   density - cluster  rotational - translational blocking has been considered  \cite{Tama:2000,Demerdash:2012}.  Alternative approaches have been proposed for the flexibility analysis of hinges in proteins using bioinformatics \cite{hingeatlas}, graph theory \cite{hingeprot,stonehinge,flexprot} and energetics \cite{flexoracle}.   Moreover, low quality experimental data due to collection conditions and  structural refinement procedures may also contribute to poor flexibility predictions.

From observation of the relationship between flexibility and local packing density, Halle \cite{Halle:2002} proposed a much simplified model called local density model (LDM), and bypassed the whole eigenmode analysis for protein B-factor prediction. In this method, the inverse of contact density, defined as the number of noncovalent neighbor atoms within a local region, is found to be proportional to atomic mean-square displacements, thus can be directly used to predict the experimental B-factors. Another interesting method is the local contact model (LCM) proposed by Zhang et al \cite{FLZhang:2002}. In this approach, the generalized order parameter of the atom is approximated by the summation of a series of exponential functions of atomic distances. Both LDM and LCM demonstrate great potential for protein flexibility prediction. Based on these approaches, many modifications have been proposed in the literature \cite{CPLin:2008,SWHuang:2008,DWLi:2009}. Among them, the weighted contact number (WCN)  is able to deliver a better accuracy than GNM using an inverse square distance function \cite{CPLin:2008}.

Recently, we have proposed a few matrix-decomposition-free methods for flexibility analysis, including  molecular nonlinear dynamics \cite{KLXia:2014b},  stochastic dynamics \cite{KLXia:2013f} and flexibility-rigidity index  (FRI)  \cite{KLXia:2013d,Opron:2014}. Among them,  flexibility-rigidity index  (FRI) has been introduced to evaluate protein flexibility and rigidity, that are further required in  a multiscale formalism called continuum elasticity with atomic rigidity (CEWAR) for  macromolecular the elasticity analysis \cite{KLXia:2013d}. The FRI method appears to be akin to  the ``flexibility index'' proposed independently by von der Lieth et al. \cite{vonderLieth:1996} and Jacobs et al. \cite{Jacobs:2001}  to describe bond strengths. However,  these flexibility indices have little in common with FRI, which does not resort to any protein interaction Hamiltonian for predicting protein  flexibility and rigidity. Instead, the FRI is a structure based approach. The fundamental assumptions of the FRI method are as follows. Protein functions, such as flexibility, rigidity, and energy, are fully determined by the structure of the protein and its environment, and the protein structure is in turn   determined by   the relevant interactions. Therefore, whenever the protein structure is available, there is no need to analyze protein flexibility and rigidity by tracing back to the protein interaction Hamiltonian. Consequently, the FRI bypasses the  ${\cal O}(N^3)$ matrix diagonalization. In fact,   FRI does not even require the 3D geometric information of the protein structure. It assesses   topological connectivity of the protein distance geometry and analyzes the geometric compactness of the protein structure. It can be regarded as a kernel generalization of the local density model \cite{Halle:2002}. Our initial FRI  \cite{KLXia:2013d} has the computational complexity of of ${\cal O}(N^2)$ and our fast FRI (fFRI) \cite{Opron:2014} based on a cell lists algorithm \cite{Allen:1987} is of ${\cal O}(N)$. The FRI and the fFRI have been extensively validated by a set of 365 proteins for parametrization, accuracy and reliability. The parameter free fFRI is about ten percent  more accurate than the GNM on the 365 protein test set and is orders of magnitude faster than GNM on a set of 44 proteins. FRI is able to predict the B-factors of an HIV virus capsid (313 236 residues) in less than 30 seconds on a single-core  processor, which would require   GNM more than 120 years  to accomplish if the computer memory is not a problem \cite{Opron:2014}.

However, earlier FRI methods do not work for many protein structures that NMA and GNM are unable to deliver good predictions. In addition to problems mentioned above,  the neglecting of  multiple characteristic length scales in  protein structures is another drawback of {present} flexibility analysis. Indeed,  biomolecules have many characteristic  length scales, ranging from  covalent bond scale, hydrogen bond scale, wan der Waals bond scale,  intraresidue scale, interresidue scale,  alpha helix  and  beta sheet scale,  domain scale and protein interaction scale.   When  GNM or FRI  is parametrized at a given cutoff or scale parameter, it captures only a subset  of the characteristic length scales but inevitably misses other characteristic length scales of the protein. Consequently, these methods fail to offer accurate B-factor prediction for many multiscale biomolecules. A multiscale  strategy has been proposed to resolve this problem by introducing two or three kernels that are parametrized at relatively small, medium and/or large length scales in the FRI formulation \cite{Opron:2015a}. We demonstrate that the resulting multiscale FRI (mFRI) works extremely well for many proteins that the GNM method fails to offer accurate flexibility analysis \cite{Opron:2015a}. Based on a set of 364 proteins, mFRI  is 20\% more accurate than GNM. It is interesting to note that there is no obvious way to incorporate multiple length scales in the aforementioned matrix diagonalization based approaches.

Although being developed independently, our FRI methods are akin to  LDM,  LCM and WCN in terms of  matrix diagonalization free. However, our FRI methods differ from LDM, LCM and WCN in the following  aspects. First, our original FRI methods were motivated from continuum mechanics, the CEWAR  \cite{KLXia:2013d,Opron:2014}. As a result, our FRI methods offer not only discrete flexibility index and discrete rigidity index, but also continuous flexibility function and continuous rigidity function. The latter is equivalent to volumetric biomolecular density distribution. Consequently, one can use the FRI rigidity function to fit electron microscope (EM) density maps  \cite{MTopf:2008,Wriggers:1999,KLXia:2015b}. A byproduct of our FRI rigidity function is the smooth biomolecular surface extracted by setting an isosurface value, including the Gaussian surface  as a special case \cite{Krone:2012,KLXia:2015d}. In contrast,  LDM, LCD and WCN do not admit any continuum representation.   Additionally, the discrete FRI formulations differ from those of LDM, LCD and WCN by admitting the diagonal term in the summation.  Moreover, we have considered the multiscale effects in biomolecules. Our mFRI captures biomolcular    thermal fluctuations  at various  length scales and thus substantially improves  the accuracy of the original FRI method. Finally, we have proposed  an anisotropic FRI  (aFRI) method to describe biomolecular collective motions.  A unique feature of our  aFRI method is that it allows adaptive Hessian matrices, from a completely global $3N\times 3N$ matrix to completely local  $3 \times 3 $ matrices. Therefore, one can use aFRI to pinpoint one's flexibility analysis to a given domain or region.

The objective of the present work is to develop FRI methods for the flexibility analysis of protein-nucleic acid complexes. Protein and nucleic acid  are dramatically different biomolecules. Amino acid residues and {nucleotides}  have different length scales and interaction characteristics. Therefore, a good model should  not only allow residues and/or {nucleotides} to be treated with different length scales, but also adapt a multiscale description of each residue and/or {nucleotide}. Unlike elastic network models that are parametrized in only one length scale for each particle,  the mFRI provides a simultaneous multiscale description. Therefore, the present mFRI is able to better capture multiscale collective motions of protein-nucleic acid complexes. Additionally, many protein-nucleic acid complexes are very large biomolecules and pose difficulty to conventional mode decomposition based methods. The  ${\cal O}(N)$ scaling FRI methods provide an efficient approach to the flexibility analysis of  large protein-nucleic acid complexes.

 The rest of this paper is organized as follows. Section \ref{sec:methods} is devoted to  methods and algorithms.  To establish notation and facilitate further discussion, the basic FRI approach is briefly discussed. We then present the multikernel based  mFRI method to improve the accuracy of biomolecular
flexibility analysis. The basic formulation of the aFRI is discussed.  In Section \ref{sec:validation}, we first analyze the benefit of adding an additional kernel with an appropriate length scale by comparing the performance of the B-factor prediction for a set of 64 protein-nucleic acid complexes introduced by Yang et al. \cite{Yang:2006} and  a larger database of 203 high resolution protein-nucleic acid structures.
Three different coarse-grain representations of    protein-nucleic acid complexes introduced by Yang et al \cite{Yang:2006} are examined. 
 Section \ref{Sec:CaseStudy} is devoted to the application of the mFRI and aFRI methods. We consider a  large ribosomeal structure to explore the utility and demonstrate the performance of the proposed mFRI. Further, we explore  the use of aFRI for the  prediction of collective motions of bridge helix, trigger loop and nucleic acids in an RNA polymerase.

\section{Methods and algorithms}\label{sec:methods}

\subsection{Flexibility-rigidity index}\label{sec:Flexibility}

In FRI, the topological connectivity of a biomolecule is measured by rigidity index and flexibility index. In particular, the rigidity index represents the protein density profile. Consider an  $N$-atom representation of a biomolecule. The coordinates of these atoms are given as $\{ {\bf r}_{j}| {\bf r}_{j}\in \mathbb{R}^{3}, j=1,2,\cdots, N\}$. We denote $  \|{\bf r}_i-{\bf r}_j\|$ the Euclidean space distance between the $i$th   atom  and the $j$th  atom. A general correlation kernel, $ \Phi( \|{\bf r} - {\bf r}_j \|;\eta_{j})$, is  a  real-valued monotonically decreasing radial basis function satisfying
\begin{eqnarray}\label{eq:couple_matrix1-1}
\Phi( \|{\bf r} - {\bf r}_j \|;\eta_{j})&=&1 \quad {\rm as }\quad  \|{\bf r} - {\bf r}_j \| \rightarrow 0\\\
\Phi( \|{\bf r} - {\bf r}_j \|;\eta_{j})&=&0 \quad {\rm as }\quad  \|{\bf r} - {\bf r}_j \| \rightarrow\infty,
\end{eqnarray}
where $\eta_{j}$ is an atomic type dependent scale parameter.
The correlation between the $i$th and $j$th particles is given by
\begin{eqnarray}\label{eq:couple_matrix0}
{C}_{ij} =  \Phi( \|{\bf r}_i - {\bf r}_j \|;\eta_{j}).
\end{eqnarray}
The correlation matrix $\{ C_{ij}\}$ can be computed to visualize the connectivity among protein particles.

We   define a position (${\bf r}$)  dependent  (continuous)  rigidity function or density function  \cite{KLXia:2013d,Opron:2014}
\begin{eqnarray}\label{eq:rigidity3}
 \mu({\bf r}) & = & \sum_{j=1}^N w_{j} \Phi( \|{\bf r} - {\bf  r}_j \|;\eta_{j} ),
 \end{eqnarray}
 where $w_{j}$ is an atom type dependent weight. For example,  carbon,  nitrogen and  phosphorus atoms can have different weights. Equation (\ref{eq:rigidity3}) can be understood as a discrete to continuum mapping. It maps a set of discrete values $\{w_j\}$ at $\{ {\bf  r}_j\}$ to the continuum domain. Although  Delta sequences of the positive type discussed in an earlier work  \cite{GWei:2000} are all good choices,   generalized exponential  functions
\begin{eqnarray}\label{eq:couple_matrix1}
\Phi(\|{\bf r} - {\bf r}_j \|;\eta_{j}) =    e^{-\left(\|{\bf r} - {\bf r}_j \|/\eta_{j}\right)^\kappa},    \quad \kappa >0
\end{eqnarray}
and  generalized Lorentz functions
\begin{eqnarray}\label{eq:couple_matrix2}
 \Phi(\|{\bf r} - {\bf r}_j \|;\eta_{j}) = \frac{1}{1+ \left( \|{\bf r} - {\bf r}_j \|/\eta_{j}\right)^{\upsilon}},  \quad  \upsilon >0
 \end{eqnarray}
have been commonly used in our recent work    \cite{KLXia:2013d,Opron:2014,Opron:2015a}.
Since the  rigidity function can be directly interpreted as a density distribution, it can been used to define the rigidity surface of a biomolecule by taking an isovalue.  By taking $\kappa =2$
in Eq. (\ref{eq:couple_matrix1}), we result in a formula for a Gaussian surface from Eq. (\ref{eq:rigidity3}).

Similarly, we define a position (${\bf r}$)  dependent (continuous) flexibility function  \cite{KLXia:2013d,Opron:2014}
\begin{eqnarray}\label{eq:flexibility1}
 F({\bf r}) & = & \frac{1}{\sum_{j=1}^N w_{j} \Phi( \|{\bf r} - {\bf  r}_j \|;\eta_{j} )}.
 \end{eqnarray}
This function is well defined in the computational domain containing the biomolecule. The flexibility function can be visualized by its projection on a given surface, such as the solvent excluded surface of a biomolecule.

The (discrete) rigidity index for the $i$th particle is obtained by restricting ${\bf r}$ to a given atomic position ${\bf r}_i$
\begin{eqnarray}\label{eq:rigidity4}
 \mu_i & = & \sum_{j=1}^N w_{j} \Phi( \|{\bf r}_i - {\bf  r}_j \|;\eta_{j} ).
 \end{eqnarray}
Here $ \mu_i$ measures the total density or rigidity at the $i$th particle. In a similar manner, we define a set of (discrete) flexibility indices by
\begin{eqnarray}\label{eq:flexibility2}
 f_i & = & \frac{1}{\sum_{j=1}^N w_{j} \Phi( \|{\bf r}_i - {\bf  r}_j \|;\eta_{j} )}.
 \end{eqnarray}
 The flexibility index $ f_i$ is directly associated with  the  B-factor  of $i$th particle
\begin{eqnarray}\label{eq:regression}
 B_i^t = a f_i + b, \quad \forall i =1,2,\cdots,N
\end{eqnarray}
where $ \{B_i^t\}$ are  theoretically predicted B-factors,  and $a$ and $b$ are two  constants to be determined by a simple linear regression. This allows us to use  experimental  data to validate the FRI method. In our earlier work  \cite{KLXia:2013d,Opron:2014,Opron:2015a}, we set $w_j=1$ for the coarse-grained C$_\alpha$  representation of proteins. We have also developed parameter free  FRI (pfFRI), such as $(\kappa=1, \eta=3)$ and $(\upsilon=3, \eta=3)$, to make our FRI robust for protein  C$_\alpha$ B-factor prediction.

\subsection{Multiscale Flexibility-rigidity index}\label{sec:MFRI}

The basic idea of multiscale FRI or multi-kernel FRI (mFRI) is quite simple. Since macromolecules are inherently multiscale in nature,
we utilize multiple   kernels that are parametrized at multiple length scales to  characterize the multiscale  thermal fluctuations of  macromolecules
\begin{eqnarray}\label{eq:flexibility3}
 f^{n}_i & = & \frac{1}{\sum_{j=1}^N w^{n}_{j} \Phi^{n}( \|{\bf r}_i - {\bf  r}_j \|;\eta^{n}_{j} )},
 \end{eqnarray}
where  $w^{n}_{j}$, $\Phi^{n}( \|{\bf r}_i - {\bf  r}_j \|;\eta^{n}_{j}) $ and $\eta^{n}_{j}$ are the corresponding quantities associated with the $n$th kernel.
We seek the minimization of the form
\begin{eqnarray}\label{eq:regression2}
{\rm Min}_{a^{n},b} \left\{ \sum_i \left| \sum_{n}a^n f^{n}_i + b-B^e_i\right|^2\right\}
\end{eqnarray}
where $\{B^e_i\}$ are the experimental B-factors. In principle, all parameters can be optimized. For simplicity and computational efficiency, we only determine $\{a^n\}$ and $b$ in the above  minimization process. For each kernel $\Phi^n$,  $w^n_j$  and $\eta^n_j$ will be selected according to the type of particles.

Specifically, for a simple C$_\alpha$ network, we can set  $w^n_j=1$, $\eta_j^{n}=\eta^{n}$ and choose a single kernel function parametrized at different scales. The predicted B-factors can be expressed as
\begin{eqnarray}\label{eq:flexibility4}
 B^{\rm mFRI}_i  = b+ \sum_{n=1}\frac{a^n}{\sum_{j=1}^N  \Phi( \|{\bf r}_i - {\bf  r}_j \|;\eta^{n} )}.
 \end{eqnarray}
The difference between Eqs. (\ref{eq:flexibility3}) and (\ref{eq:flexibility4}) is that, in Eqs. (\ref{eq:flexibility3}), both the kernel and the scale can be changed for different $n$. In contrast, in Eq.  (\ref{eq:flexibility4}), only the scale is changed. One can use a given kernel, such as
\begin{eqnarray}\label{eq:couple_matrixn}
 \Phi(\|{\bf r} - {\bf r}_j \|;\eta^n) = \frac{1}{1+ \left( \|{\bf r} - {\bf r}_j \|/\eta^n\right)^{3}},
 \end{eqnarray}
to achieve good multiscale predictions \cite{Opron:2015a}.

\subsection{Anisotropic flexibility-rigidity index}\label{sec:AnisoFlexibility}

 The anisotropic flexibility-rigidity index (aFRI) model was built in a very unique manner. Different from the existing normal mode analysis or anisotropic elastic network models, in which the Hessian matrix is always global, our aFRI model delivers a local and adaptive Hessian matrix. This means that for a molecule with $N$ particles, the Hessian matrix is always $3N\times 3N$ for ANM, whereas, our Hessian matrix may vary from a set of $N$ $3\times 3$ matrices for a completely local aFRI to $3N\times 3N$ for a complete global aFRI, depending on the need of a physical problem. For instance, if one is particularly interested in certain structures like alpha helices, domains, or binding sites of a protein, or certain subunits of a biomolecular complex, one design and an aFRI for these portions of the molecule. We   partition all the $N$ particles in a molecule or a biomolecular complex into a total of $M$ clusters $\{c_1, c_2,\cdots, c_k, \cdots, c_M \}$. Cluster $c_k$ has $N_k$ particles or atoms so that $N=\sum_{k=1}^M N_{k}$. We choose  clusters based on our physical interest as mentioned above. In this way, two very special situations can be found. The first one corresponds to the completely local situation, i.e.,  $N$ clusters and each cluster has only one atom. The other situation contains only one cluster, which is then completely global. It is straightforward to construct a $3N\times 3N$ Hessian matrix and analyze the collective motion. The problem arises when we consider the global motion of a selected cluster, at the same time include the influence from the rest clusters.  The essential idea is to construct a cluster Hessian matrix for each cluster individually and then incorporate the information from nearby clusters into its diagonal terms.

For example, if we want to know the thermal fluctuation of a particular cluster $c_k$ with  $N_{k}$ particles or atoms, we need to find $3N_k$ eigenvectors for the cluster.  Let us keep in mind that each position vector in $\mathbb{R}^3$ has three components, i.e., ${\bf r}=(x,y,z)$.
For each given pair of particles $i$ and $j$, we can define a local anisotropic matrix $\Phi^{ij}=\left( \Phi^{ij}_{uv} \right)$ as
\begin{equation}
\Phi^{ij}=\left(
\begin{array}{ccc}
\Phi^{ij}_{xx} & \Phi^{ij}_{xy}& \Phi^{ij}_{xz}\\
\Phi^{ij}_{yx} & \Phi^{ij}_{yy}& \Phi^{ij}_{yz}\\
\Phi^{ij}_{zx} & \Phi^{ij}_{zy}& \Phi^{ij}_{zz}
\end{array}
\right),
\end{equation}
where  $\Phi^{ij}_{uv}$ are defined as
\begin{eqnarray}\label{eq:Anisorigidity1}
 \Phi^{ij}_{uv}  = \frac{\partial}{\partial u_i} \frac{\partial}{\partial v_j} \Phi( \|{\bf r}_i - {\bf  r}_j \|; \eta_{j} ), \quad  u,v= x, y, z; i,j =1,2,\cdots,N.
\end{eqnarray}

Due to the inner connection between rigidity and flexibility, we have two different aFRI algorithms. The specially designed cluster Hessian matrix with a  smaller size can incorporate nonlocal geometric impact
and predict collective thermal motions of the cluster. The details are presented below.

\subsubsection{Anisotropic rigidity}\label{sec:arFRI}

In anisotropic rigidity based aFRI, a rigidity Hessian matrix is needed. For a cluster $c_k$, if we denote its  rigidity Hessian matrix as $ \left(\mu_{uv}^{ij}(c_k)\right)$ with elements,
\begin{eqnarray}\label{eq:Anisorigidity2}
\mu^{ij}_{uv}(c_k) =& - w_{j}\Phi^{ij}_{uv},                &\quad   i,j \in c_k; i\neq j;  u,v= x, y, z \\ \label{eq:Anisorigidity3}
\mu^{ii}_{uv}(c_k)=&  \sum_{j=1}^N w_{j} \Phi^{ij}_{uv},  &\quad   i \in c_k;  u,v= x, y, z \\ \label{eq:Anisorigidity4}
\mu^{ij}_{uv}(c_k)=&  0,                                   &\quad   i,j \notin  c_k; u,v= x, y, z.
\end{eqnarray}
In this way, the rigidity Hessian matrix is of  $3N_k\times 3N_k$ dimensions. More importantly, the information from all other clusters are built into diagonal terms, even if the cluster itself is completely localized, i.e.,  $N_k=1,~ \forall k$.

For B-factor prediction, we define a set of anisotropic rigidity (AR) based flexibility indices by
\begin{eqnarray}\label{eq:Anisorigidity34}
f_{i}^{\rm AR} =  \frac{1}{\mu^{i}_{\rm diag}},
\end{eqnarray}
where the $i$th diagonal term $\mu^{i}_{\rm diag}$ is of the form,
\begin{eqnarray}\label{eq:Anisorigidity33}
\mu^{i}_{\rm diag} &=& {\rm Tr}\left(\mu_{uv}^{i}\right)\\
                    &=&      \sum_{j=1}^N w_{j} \left[\Phi^{ij}_{xx}+\Phi^{ij}_{yy}+   \Phi^{ij}_{zz}\right].
\end{eqnarray}
Here, $f_{i}^{\rm AR}$ is employed in the linear regression  to determine B-factors.


\begin{figure}
\begin{center}
\begin{tabular}{cc}
\includegraphics[width=0.4\textwidth]{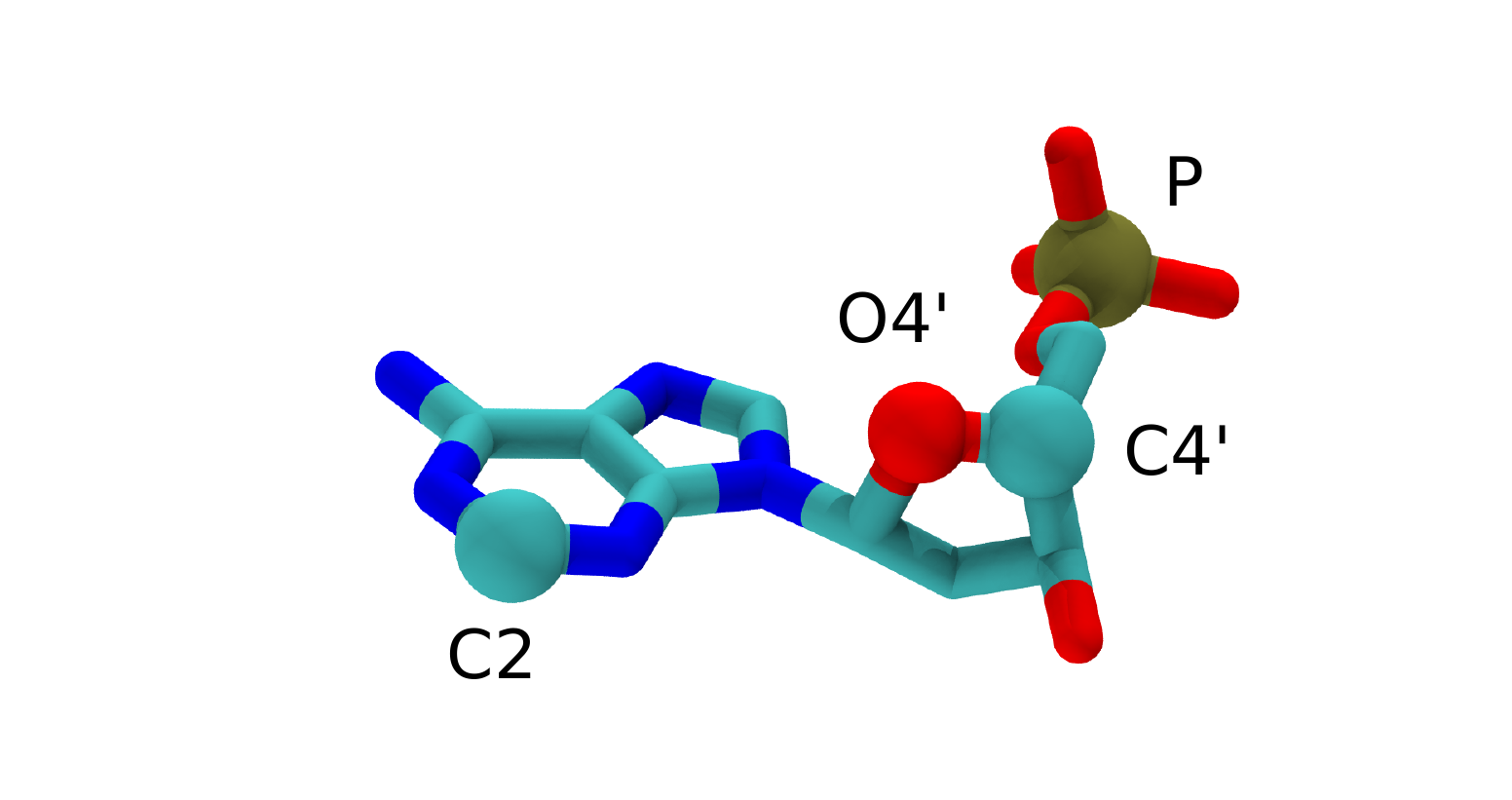}
\includegraphics[width=0.4\textwidth]{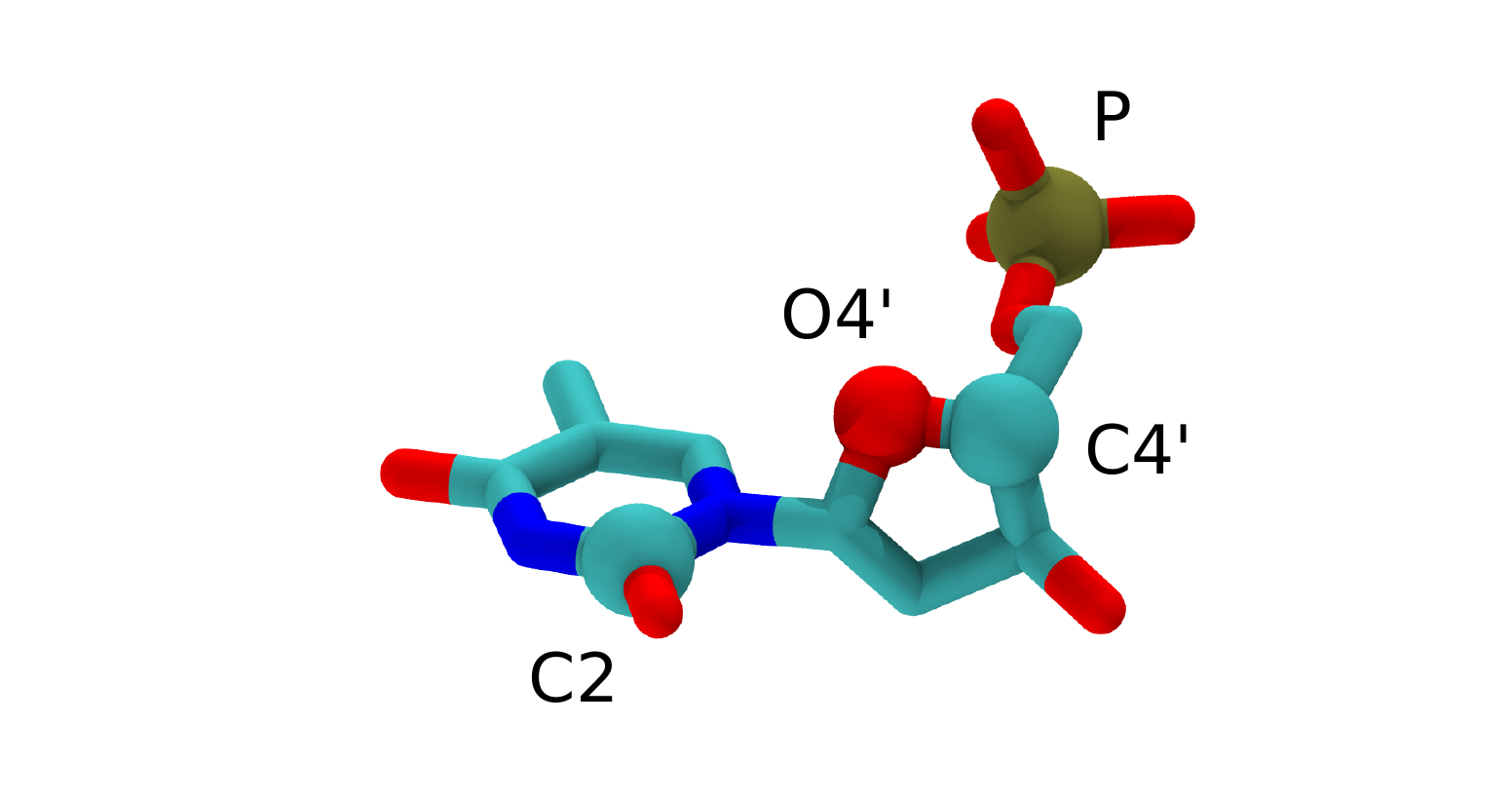}
\end{tabular}
\end{center}

\caption{  Illustration highlighting atoms used for coarse-grained representations in protein-nucleic acid complexes for FRI and GNM. In addition to protein  C$\alpha$ atoms, Model M1 considers the backbone P
atoms for nucleotides. Model M2 includes M1 atoms and adds the sugar O4' atoms for nucleotides. Model M3 includes M1 atoms and adds the sugar C4' atoms and the base C2 atoms for nucleotides.
}
\label{CoarseG}
\end{figure}

\subsubsection{Anisotropic flexibility}

The other way to construct aFRI is to construct  a flexibility Hessian  matrix, which is denoted as ${\bf F}(c_k)$ for cluster $c_k$ with elements,
\begin{eqnarray}\label{eq:Anisoflexibility}
{\bf F}^{ij}(c_k)     =&  - \frac{1}{w_{j}} (\Phi^{ij})^{-1},                &\quad   i,j \in c_k; i\neq j;  u,v= x, y, z \\ \label{eq:Anisoflexibilityy3}
{\bf F}^{ii}(c_k)=&   \sum_{j=1}^N \frac{1}{w_{j}} (\Phi^{ij})^{-1},  &\quad   i \in c_k;  u,v= x, y, z \\ \label{eq:Anisoflexibility4}
{\bf F}^{ij}(c_k)=&  0,                                     &\quad   i,j \notin  c_k; u,v= x, y, z.
\end{eqnarray}
Note that $(\Phi^{ij})^{-1}$ represents the unscaled inverse of matrix $\Phi^{ij}$ such that $\Phi^{ij}(\Phi^{ij})^{-1}=| \Phi^{ij}|$. The diagonalization of ${\bf F}(c_k)$ gives rise to eigenmodes, which represent the cluster  motions.  Additionally,  the diagonal part ${\bf F}^{ii}(c_k)$ has built in information from all particles in the system. In this way, we deliver a cluster Hessian matrix. By diagonalizing ${\bf F}(c_k)$, we obtain $3N_k$ eigenvectors for the $N_k$ particles in the cluster $c_k$ of interest.  Furthermore, instead of predicting the B-factors via the eigenvalues and eigenmodes, we   directly predict the B-factors by using our anisotropic flexibility (AF) based flexibility indices defined as,
\begin{eqnarray}\label{eq:Anisoflexibility2}
f_i^{\rm AF} &=&{\rm Tr} \left({\bf F}(c_k)\right)^{ii},   \\
                &=&  \left({\bf F}(c_k)\right)^{ii}_{xx}+ \left({\bf F}(c_k)\right)^{ii}_{yy}+ \left({\bf F}(c_k)\right)^{ii}_{zz}.
\end{eqnarray}
Finally, we employ $f_i^{\rm AF}$ to predict B-factors.

\section{ Implementation and validation}\label{sec:validation}

In this section, we parametrize and test the previously described mFRI on protein-nucleic acid structures. A immediate concern is whether the proposed mFRI is as efficient on protein-nucleic structures as it is on protein-only structures as shown in a previous study \cite{KLXia:2015a}. The accuracy of the mFRI method is tested by the  B-factor prediction of two sets of protein-nucleic acid structures, including a set of 64 molecules used in a recent GNM study  \cite{Yang:2006}  and a set of 203 molecules for more accurate parametrization of mFRI.

\newcolumntype{L}[1]{>{\raggedright\let\newline\\\arraybackslash\hspace{0pt}}m{#1}}
\newcolumntype{C}[1]{>{\centering\let\newline\\\arraybackslash\hspace{0pt}}m{#1}}
\newcolumntype{R}[1]{>{\raggedleft\let\newline\\\arraybackslash\hspace{0pt}}m{#1}}
\begin{figure}
\begin{center}
\begin{tabular}{c}
\includegraphics[width=0.45\textwidth]{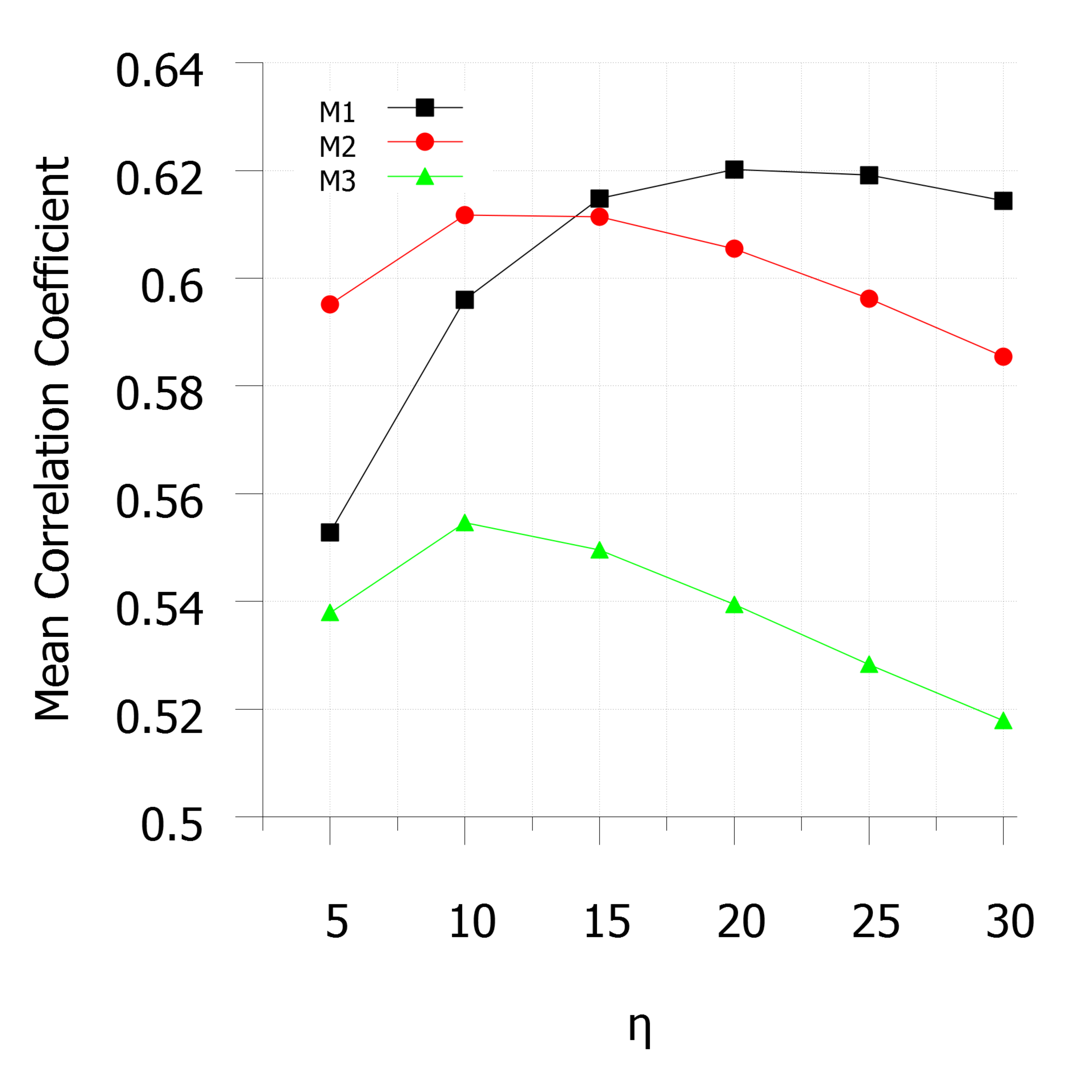}
\includegraphics[width=0.45\textwidth]{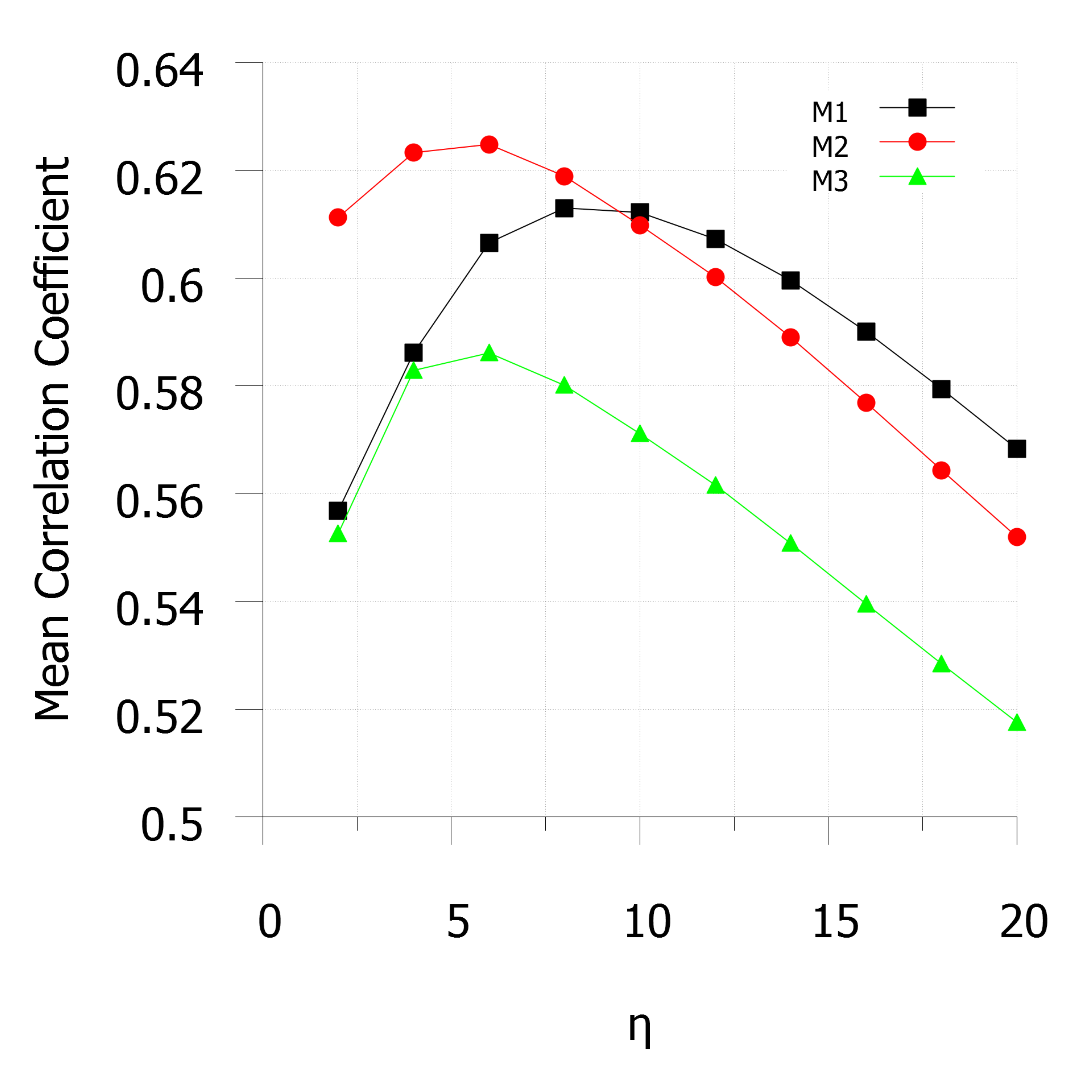}
\end{tabular}
\end{center}
\caption{ MCCs for single kernel parameter test using the M1 (squares), M2 (circles) and M3  (triangles) representations. Lorentz kernel with $\upsilon=3$ is used. The parameter $\eta$ is varied to find the maximum MCC on the test set of structures. The results for a set of 64 protein-nucleic structures   ( PDB IDs listed in Table \ref{table:64set}) 
are shown on the left, while results for a separate set of 203 structures (PDB IDs listed in Table \ref{table:203set}) is shown on the right for more general selections.}
\label{single_etasigma}
\end{figure}

\subsection{Coarse-grained representations of protein-nucleic acid complexes}\label{sec:nucleic}
In this section, we consider flexibility analysis of protein-nucleic acid complexes.  To this end, we need  coarse-grained representations. We consider three coarse-grained representation of nucleic acids to be used in conjugation with the C$\alpha$-only representation used for proteins. These three models are identical to those used by Yang et al. \cite{Yang:2006} and are named M1, M2 and M3. Model M1 consists of  the backbone P atoms and protein C$\alpha$atoms. Model M2 contains the same atoms as M1 but also includes sugar O4' atoms. Model M3 includes atoms from M1 and adds the sugar C4' atoms and base C2 atoms, see Fig. \ref{CoarseG}.

Model M1 is similar to protein C$\alpha$ representations because they are both backbone-only representations. The atoms in M1 are 6 bonds apart while C$\alpha$ atoms are 3 bonds apart. Model M2 includes P atoms and adds the O4' atoms located on the ribose portion of the nucleotide. Finally, model M3 includes atoms of P, C4' and  C2, a carbon from the base portion of the nucleotide, see Fig. \ref{CoarseG}. As point out by Yang et al. \cite{Yang:2006}, nucleotides are approximately three times more massive than amino acids and so model M3 with three nodes per nucleotide is consistent in this sense with using C$\alpha$ atoms for the protein representation.

\subsection{Multiscale/Multikernel FRI }\label{sec:multiFRI}

To parametrize and test the accuracy of multikernel fFRI on protein-nucleic acid structures, we use a dataset from Yang et al. \cite{Yang:2006} containing 64 structures. In addition, we construct a larger database of 203 high resolution structures. This expanded protein-nucleic structure set was obtained by searching the Protein Data Bank (PDB) for structures that contain both Protein and DNA and structure which have an X-ray resolution between 0.0 and 1.75 \AA. All PDB files are processed by removing low occupancy atomic coordinates for structures having residues with multiple possible coordinates.  The PDB IDs of the 64 and 203 structures can be found in Table \ref{table:64set} and Table \ref{table:203set}, respectively.

To quantitatively  assess the performance of the proposed multikernel FRI method, we  consider the correlation coefficient (CC)
\begin{eqnarray}\label{correlation}
   {\rm CC}=\frac{\sum^N_{i=1}\left(B^e_i-\bar{B}^e \right)\left( B^t_i-\bar{B}^t \right)}
   { \left[\sum^N_{i=1}(B^e_i- \bar{B}^e)^2\sum^N_{i=1}(B^t_i-\bar{B}^t)^2\right]^{1/2}},
\end{eqnarray}
where $\{B^t_i,  i=1,2,\cdots,N\}$ are a set of predicted B-factors by using the proposed method and $\{B^e_i, i=1,2,\cdots, N\}$ are a set of experimental B-factors read from the PDB file. Here $\bar{B}^t$ and $\bar{B}^e$ the statistical averages of theoretical and experimental B-factors, respectively.

\subsubsection{Multikernel FRI testing on protein-nucleic structures }\label{sec:multisearch}

Previous tests of single kernel FRI indicate that the Lorentz type and exponential type correlation kernels are the two most accurate kernel types. This leads us to try the combination of  these two types of kernels. The resulting multikernel FRI method requires four parameters, namely, $\kappa$ and $\eta$ for the exponential kernel and $\upsilon$ and $\eta$ for the Lorentz kernel.

\subsubsection{Single kernel FRI testing }\label{sec:singleFRI}

In order to compare FRI and   GNM methods for protein-nucleic acid structures, we test our single kernel FRI at a range of $\eta$ values. For this test we use the Lorentz kernel with $\upsilon=3$ for B-factor prediction on both structures sets and all three representations (M1, M2 and M3). The results are shown in Figure \ref{single_etasigma}. For the 64 structure set, single kernel FRI has a maximum mean correlation coefficient (MCC) to experimental B-factors for M1, M2 and M3 representations of 0.620, 0.612 and 0.555. Comparatively, GNM had a MCC of approximately 0.59, 0.58 and 0.55 for M1, M2 and M3 for the same data set \cite{Yang:2006}. The maximum MCCs for FRI on the larger data set for M1, M2 and M3 are 0.613, 0.625 and 0.586, respectively. The M1 and M2 representations perform better than the M3 representation.

\subsubsection{Parameter-free multikernel FRI }\label{sec:pfFRI}

\begin{figure}
\begin{center}
\begin{tabular}{ccc}
\includegraphics[width=0.33\textwidth]{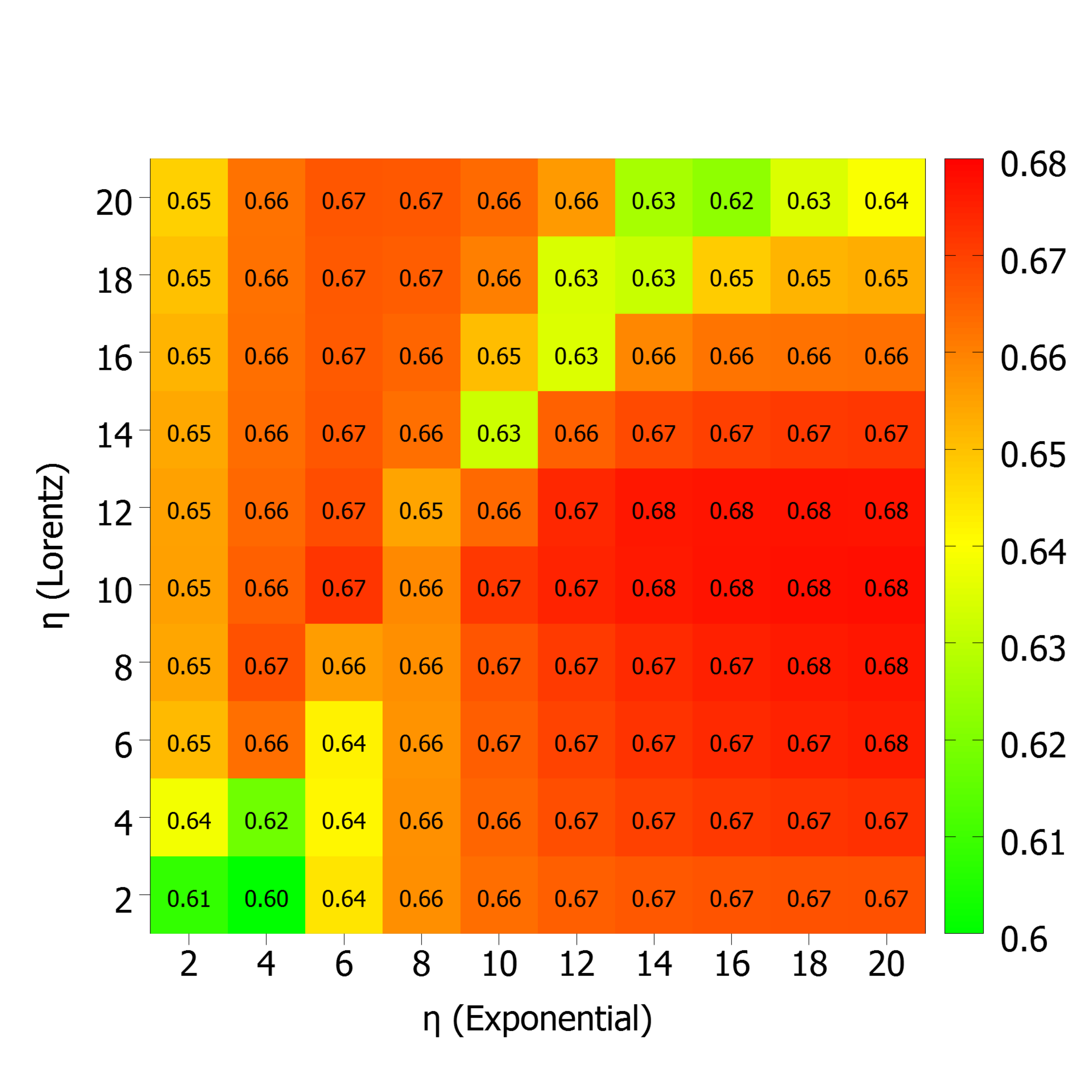}
\includegraphics[width=0.33\textwidth]{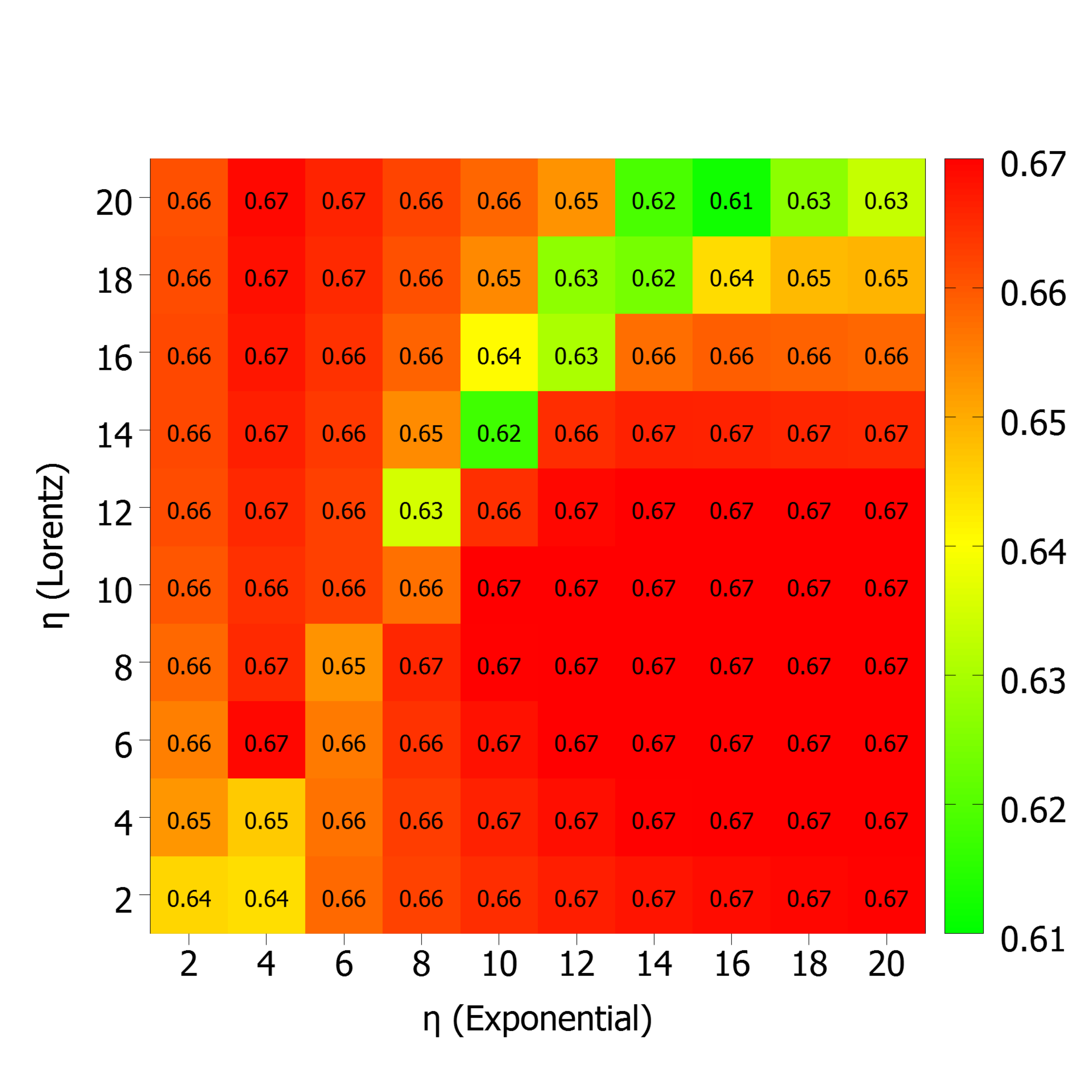}
\includegraphics[width=0.33\textwidth]{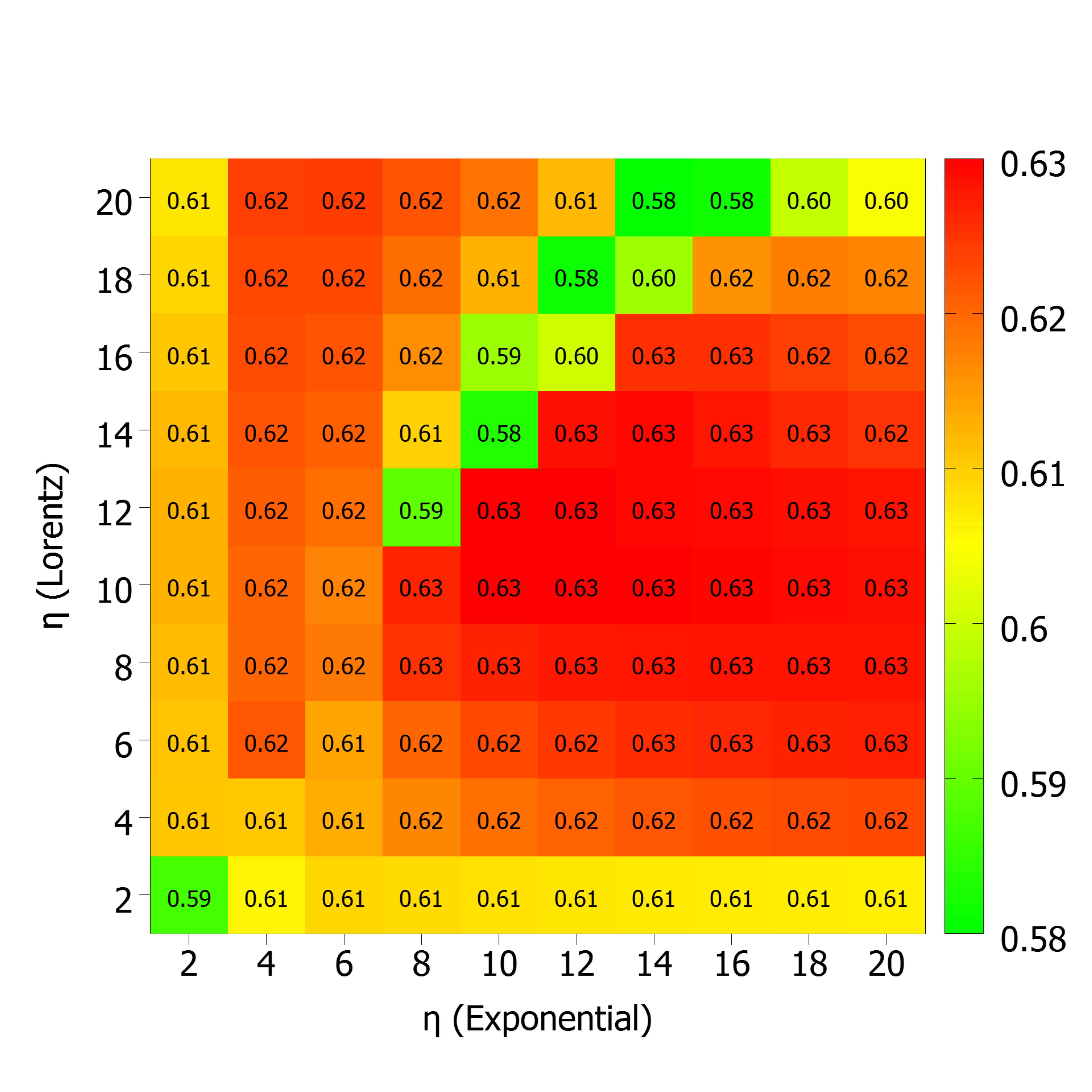}
\end{tabular}
\end{center}
\caption{ Mean correlation coefficients (MCCs) for two-kernel FRI models on a set of 203 protein-nucleic structures. From left to right, MCC values are shown for M1, M2 and M3 representations. We use one Lorentz kernel with $\upsilon=3.0$ and one exponential kernel with $\kappa=1.0$.   The values of parameter $\eta$ for both kernels are varied from 2 to 20 \AA.
}
\label{multi_etasigma}
\end{figure}

As with protein-only structures, we develop   multikernel FRIs  with multiple kernels to improve accuracy of prediction on protein-nucleic acid structures. In order to simplify the FRI method, we try to develop an accurate parameter-free version  for a two-kernel mFRI. We use a combination of one Lorentz   and one exponential kernel. Values for parameters $\upsilon$ and $\kappa$ are set  to 3.0 and 1.0 respectively based on the results of previous FRI studies \cite{Opron:2014}. The optimal values for $\eta$ in both kernels are determined by testing a range of possible values from 2 to 20 \AA. All three  representations (M1, M2 and M3) described previously are considered. The results of these tests on the set of 203 protein-nucleic acid structures are shown in Figure \ref{multi_etasigma}.

\begin{table}
\caption{  Correlation coefficients (CCs) between predicted and experimental B-factors for the set of 64 protein-nucleic structures \cite{Yang:2006}. Here N1, N2 and N3 values represent the number of atoms used for the M1, M2 or M3 representations for each structure.  We use the parameter-free two-kernel mFRI model, i.e., one exponential kernel ($\kappa=1$ and $\eta=18$ \AA) and one Lorentz kernels ($\upsilon=3$, $\eta=18$ \AA. {PDB IDs marked with an asterisk (*) indicate structure containing only nucleic-acid residues. }}
\begin{center}
\begin{adjustbox}{width=\textwidth,totalheight=0.93\textheight,keepaspectratio}
\begin{tabular}{lcrcrcr}
      & \multicolumn{2}{c}{\textbf{M1}} & \multicolumn{2}{c}{\textbf{M2}} & \multicolumn{2}{c}{\textbf{M3}} \bigstrut[b]\\
\hline
\hline
PDB ID & CC & \multicolumn{1}{c}{N1} & CC & \multicolumn{1}{c}{N2} & CC & \multicolumn{1}{c}{N3} \bigstrut\\
\hline
1asy  & 0.647 & 1114  & 0.645 & 1248  & 0.631 & 1382 \bigstrut[t]\\
1b23  & 0.751 & 471   & 0.774 & 537   & 0.714 & 603 \\
1c0a  & 0.763 & 653   & 0.704 & 721   & 0.598 & 789 \\
1CX0  & 0.821 & 162   & 0.763 & 234   & 0.627 & 306 \\
1drz  & 0.846 & 162   & 0.754 & 234   & 0.585 & 306 \\
1efw  & 0.537 & 1286  & 0.647 & 1412  & 0.660 & 1538 \\
1egk*  & 0.273 & 104   & 0.298 & 212   & 0.267 & 320 \\
1ehz*  & 0.623 & 62    & 0.706 & 124   & 0.722 & 186 \\
1evv*  & 0.710 & 62    & 0.769 & 124   & 0.770 & 186 \\
1f7u  & 0.577 & 670   & 0.588 & 734   & 0.603 & 798 \\
1ffk  & 0.759 & 6482  & 0.793 & 9310  & 0.809 & 12138 \\
1ffy  & 0.520 & 991   & 0.549 & 1066  & 0.568 & 1141 \\
1fg0*  & 0.720 & 498   & 0.723 & 996   & 0.721 & 1494 \\
1fir*  & 0.687 & 61    & 0.576 & 122   & 0.439 & 183 \\
1fjg  & 0.461 & 3915  & 0.585 & 5428  & 0.600 & 6941 \\
1gid*  & 0.649 & 316   & 0.643 & 632   & 0.583 & 948 \\
1gtr  & 0.724 & 603   & 0.747 & 677   & 0.645 & 751 \\
1h3e  & 0.717 & 507   & 0.724 & 586   & 0.645 & 663 \\
1h4s  & 0.671 & 1011  & 0.704 & 1076  & 0.626 & 1141 \\
1hr2*  & 0.599 & 313   & 0.589 & 628   & 0.585 & 943 \\
1i94  & 0.489 & 3923  & 0.615 & 5437  & 0.652 & 6951 \\
1i9v*  & 0.615 & 73    & 0.631 & 147   & 0.642 & 220 \\
1j1u  & 0.730 & 372   & 0.671 & 446   & 0.456 & 520 \\
1j2b  & 0.686 & 1300  & 0.712 & 1448  & 0.672 & 1596 \\
1j5a  & 0.532 & 3158  & 0.548 & 5932  & 0.510 & 8706 \\
1j5e  & 0.427 & 3909  & 0.546 & 5422  & 0.553 & 6935 \\
1jj2  & 0.799 & 6567  & 0.839 & 9443  & 0.836 & 12319 \\
1jzx  & 0.586 & 3158  & 0.600 & 5932  & 0.561 & 8706 \\
1l8v*  & 0.700 & 312   & 0.688 & 626   & 0.672 & 940 \\
1l9a  & 0.849 & 211   & 0.789 & 336   & 0.675 & 461 \\
1lng  & 0.780 & 183   & 0.595 & 280   & 0.405 & 377 \\
1m5k  & 0.904 & 402   & 0.841 & 622   & 0.760 & 842 \\
1m5o  & 0.921 & 405   & 0.872 & 629   & 0.810 & 853 \\
1mfq  & 0.773 & 341   & 0.688 & 468   & 0.543 & 595 \\
1mms  & 0.507 & 317   & 0.548 & 433   & 0.646 & 549 \\
1n32  & 0.388 & 3916  & 0.494 & 5447  & 0.517 & 6978 \\
1nbs*  & 0.547 & 270   & 0.566 & 540   & 0.573 & 810 \\
1o0c  & 0.766 & 602   & 0.758 & 676   & 0.636 & 750 \\
1qf6  & 0.608 & 710   & 0.578 & 779   & 0.540 & 848 \\
1qrs  & 0.671 & 603   & 0.672 & 677   & 0.586 & 751 \\
1qtq  & 0.620 & 602   & 0.640 & 676   & 0.596 & 750 \\
1qu2  & 0.520 & 991   & 0.549 & 1066  & 0.568 & 1141 \\
1qu3  & 0.579 & 954   & 0.599 & 1029  & 0.613 & 1104 \\
1rc7  & 0.599 & 256   & 0.566 & 296   & 0.470 & 336 \\
1s72  & 0.823 & 6636  & 0.839 & 9507  & 0.831 & 12378 \\

1ser  & 0.748 & 855   & 0.743 & 917   & 0.657 & 978 \\
1sj3  & 0.880 & 167   & 0.805 & 240   & 0.614 & 313 \\
1tn2*  & 0.686 & 62    & 0.712 & 124   & 0.676 & 186 \\
1tra*  & 0.624 & 62    & 0.670 & 124   & 0.660 & 186 \\
1ttt  & 0.578 & 1401  & 0.564 & 1587  & 0.515 & 1773 \\
1u0b  & 0.757 & 535   & 0.754 & 609   & 0.621 & 683 \\
1u6b  & 0.476 & 312   & 0.490 & 531   & 0.506 & 750 \\
1u9s*  & 0.446 & 155   & 0.432 & 310   & 0.419 & 465 \\
1vby  & 0.877 & 167   & 0.792 & 240   & 0.587 & 313 \\
1vc0  & 0.878 & 167   & 0.804 & 240   & 0.611 & 313 \\
1vc5  & 0.861 & 164   & 0.840 & 234   & 0.685 & 304 \\
1y0q*  & 0.491 & 230   & 0.484 & 463   & 0.472 & 696 \\
1y26*  & 0.677 & 70    & 0.697 & 141   & 0.709 & 212 \\
1yfg*  & 0.565 & 64    & 0.600 & 128   & 0.623 & 192 \\
1yhq  & 0.835 & 6636  & 0.840 & 9507  & 0.831 & 12378 \\
1yij  & 0.836 & 6636  & 0.851 & 9507  & 0.842 & 12378 \\
2tra*  & 0.614 & 65    & 0.614 & 130   & 0.613 & 195 \\
3tra*  & 0.645 & 64    & 0.615 & 128   & 0.620 & 192 \\
4tra*  & 0.679 & 62    & 0.715 & 124   & 0.694 & 186 \\
\hline
\hline
\end{tabular}%
\label{table:64set}
\end{adjustbox}
\end{center}
\end{table}

\begin{table}
\caption{The PDB IDs of the 203 high resolution protein-nucleic structures used in our single-kernel FRI parameter test. IDs marked with an asterisk indicate those containing only nucleic acids residues. }
\begin{center}
\begin{tabular}{|c|c|c|c|c|c|c|c|c|c|}
\hline
\hline
PDB ID & PDB ID & PDB ID & PDB ID & PDB ID & PDB ID & PDB ID & PDB ID & PDB ID & PDB ID \bigstrut\\
\hline
1A1H  &1A1I  &1AAY  &1AZP  &1BF4  &1C8C  &1D02  &1D2I  &1DC1  &1DFM \\\hline
1DP7  &1DSZ  &1EGW  &1EON  &1F0V  &1FIU  &1H6F  &1I3W  &1JK2  &1JX4 \\\hline
1K3W  &1K3X  &1L1Z  &1L3L  &1L3S  &1L3T  &1L3V  &1LLM  &1MNN  &1NJX \\\hline
1NK0  &1NK4  &1OJ8  &1ORN  &1PFE  &1QUM  &1R2Z  &1RFF  &1RH6  &1SX5 \\\hline
1T9I  &1U4B  &1VTG  &1WTO  &1WTQ  &1WTV  &1XJV  &1XVK  &1XVN  &1XVR \\\hline
1XYI  &1ZS4  &2ADW  &2AXY  &2BCQ  &2BCR  &2BOP  &2C62  &2C7P  &2EA0 \\\hline
2ETW  &2EUW  &2EUX  &2EUZ  &2EVF  &2EVG  &2FMP  &2GB7  &2HAX  &2HEO \\\hline
2HHV  &2IBT  &2IH2  &2ITL  &2NQ9  &2O4A  &2OAA  &2ODI  &2P2R  &2PY5 \\\hline
2Q10  &2R1J  &2VLA  &2VOA  &2WBS  &2XHI  &2Z70  &2ZKD  &3BIE  &3BKZ \\\hline
3BM3  &3BS1  &3D2W  &3EY1  &3EYI  &3FC3  &3FDE  &3FDQ  &3FSI  &3FYL \\\hline
3G00  &3G9M  &3G9O  &3G9P  &3GO3  &3GOX  &3GPU  &3GQ4  &3HPO  &3HT3 \\\hline
3HTS  &3I0W  &3I2O  &3I3M  &3I49  &3I8D  &3IGK  &3JR5  &3JX7  &3JXB \\\hline
3JXY  &3JXZ  &3KDE  &3KXT  &3M4A  &3MR3  &3MXM  &3NDH  &3O1M  &3O1P \\\hline
3O1S  &3O1T  &3O1U  &3OQG  &3PV8  &3PVI  &3PX0  &3PX4  &3PX6  &3PY8 \\\hline
3QEX  &3RKQ  &3RZG  &3S57  &3S5A  &3SAU  &3SJM  &3TAN  &3TAP  &3TAQ \\\hline
3TAR  &3THV  &3TI0  &3U6E  &3U6P  &3V9W  &3ZDA  &3ZDB  &3ZDC  &3ZDD \\\hline
4A75  &4B21  &4B9S  &4DFK  &4DQI  &4DQP  &4DQQ  &4DS4  &4DS5  &4DSE \\\hline
4DSF  &4E0D  &4ECQ  &4ECV  &4ECX  &4ED0  &4ED2  &4ED7  &4ED8  &4EZ6 \\\hline
4F1H  &4F2R  &4F2S  &4F3O  &4F4K  &4F8R  &4FPV  &4GZ1  &4GZN  &4HC9 \\\hline
4HIK  &4HIM  &4HLY  &4HTU  &4HUE  &4HUF  &4HUG  &4IBU  &4IX7  &4KLG \\\hline
4KLI  &4KLM  &4KMF  \\
\hline
\hline
\end{tabular}%
\label{table:203set}
\end{center}
\end{table}

As expected, the addition of another kernels results in an overall increase in accuracy for the 203 complex set. For two-kernel mFRI, the MCCs increase up to 0.68 for M1,  0.67  for M2 and 0.63 for M3.  The choice of $\eta$ turns out to be very robust based on  results shown in Figure \ref{multi_etasigma}.

We have also carried out a similar test of two-kernel mFRI ($\upsilon=3.0$ and $\kappa=1.0$) for the set of 64 protein-nucleic acid structures. Note that this has many large complexes. The MCCs for M1, M2 and M3 models are 0.668, 0.666 and 0.620, respectively, which are similar to what we have found for the set of 203 structures. {The set of 64 structures includes 19 structures composed of nucleic acids and no amino acids. The MCCs for this nucleic acid-only subset 0.608,		0.617 and		0.603 for M1, M2 and M3 models. The correlation coefficients for all 64 individual molecular complexes are listed in Table \ref{table:64set}.}

To summarize the performance of  Gaussian network model, single kernel FRI, and    two-kernel mFRI, we list their MCCs for the 64 protein-nucleic acid structures in Table \ref{table:compare}. It can be seen that, the FRI outperforms GNM in all three representations, and two-kernel mFRI further significantly improves the accuracy of our method  and achieves up to 15\% improvement compared with GNM \cite{Yang:2006}. Based on our earlier test \cite{Opron:2015a}, we believe that  our three-kernel mFRI can  deliver a better prediction.

\begin{table}
\begin{center}
\caption{   MCCs of Gaussian network model (GNM) \cite{Yang:2006}, single kernel flexibility-rigidity index (FRI) and  two-kernel mFRI   for three coarse-grained representations (M1, M2,and  M3). A set of 64 protein-nucleic acid structures \cite{Yang:2006} is used.  }
\begin{tabular}{cccc}
     &{\textbf{GNM}}\cite{Yang:2006}  &{\textbf{FRI}}   &{\textbf{Two-kernel mFRI}}   \\
\hline
M1  & 0.59   & 0.620  & 0.666   \\\hline
M2  & 0.58   & 0.612  & 0.668   \\\hline
M3  & 0.55   & 0.555  & 0.620 \\
\hline
\end{tabular}
\label{table:compare}
\end{center}
\end{table}

\section{Applications   }\label{Sec:CaseStudy}

\newcolumntype{a}{>{\columncolor{Gray}}c}
\newcolumntype{b}{>{\columncolor{white}}c}

In this section we briefly explore  the  applications of the mFRI and aFRI methods to large protein-nucleic acid complexes. We highlight a few particular examples where mFRI improves upon previous FRI methods, in particular, for the flexibility prediction of ribosomes. Further, we show how aFRI is well suited for the study of the dynamics of large macromolecular complexes using the bacterial RNA polymerase active site as an example.

\subsection{Multikernel FRI flexibility prediction for protein-nucleic acid structures -   ribosomes}\label{sec:ribo}

\begin{figure}
\begin{center}
\subfigure[Complete ribosome with bound tRNAs PDB ID: 4V4J.]{%
\includegraphics[width=0.80\textwidth]{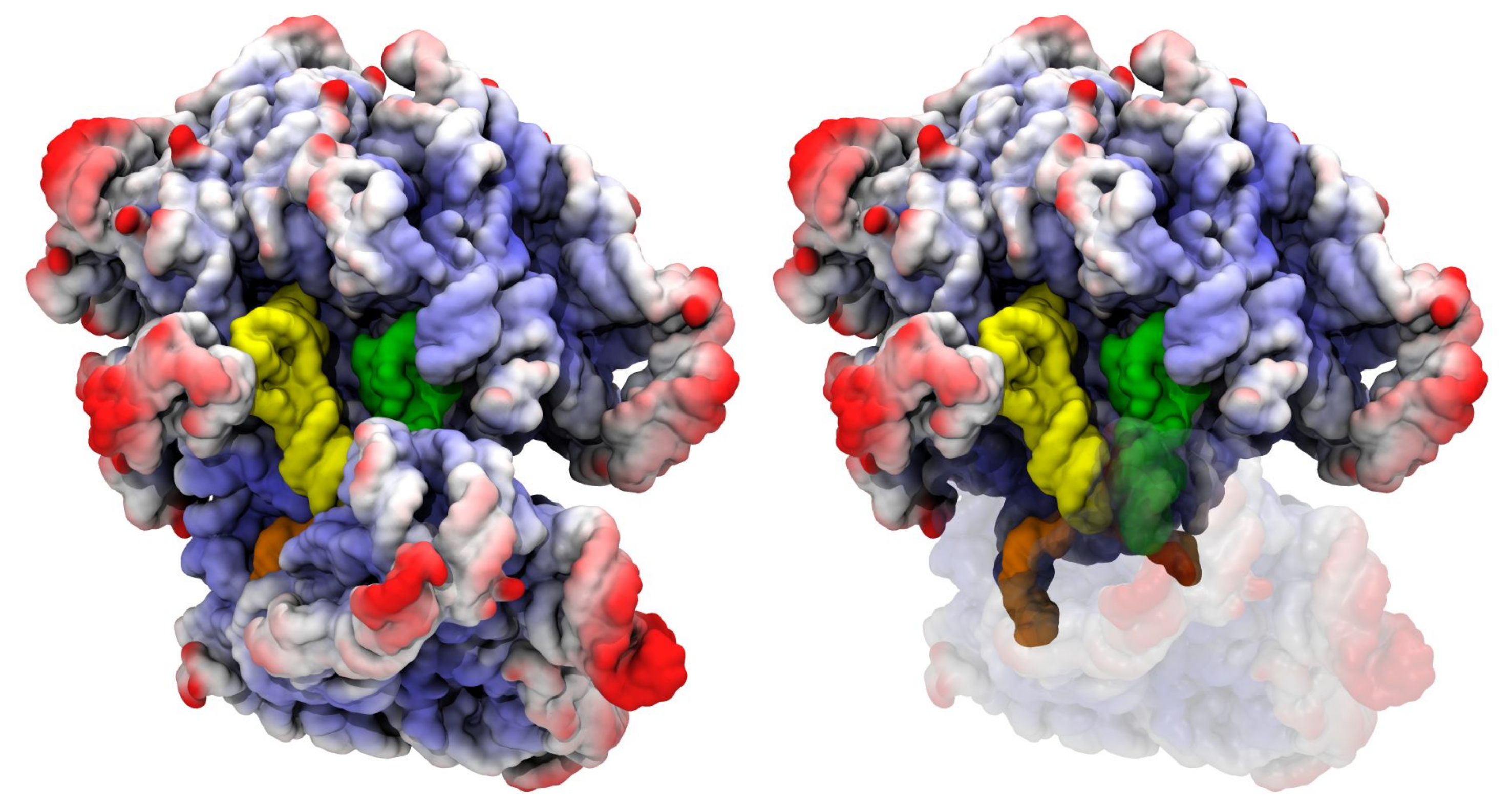}
\label{fig:subfigure1}}
\subfigure[Ribosome 50S subunit PDB ID: 1YIJ B factors]{%
\includegraphics[width=0.50\textwidth]{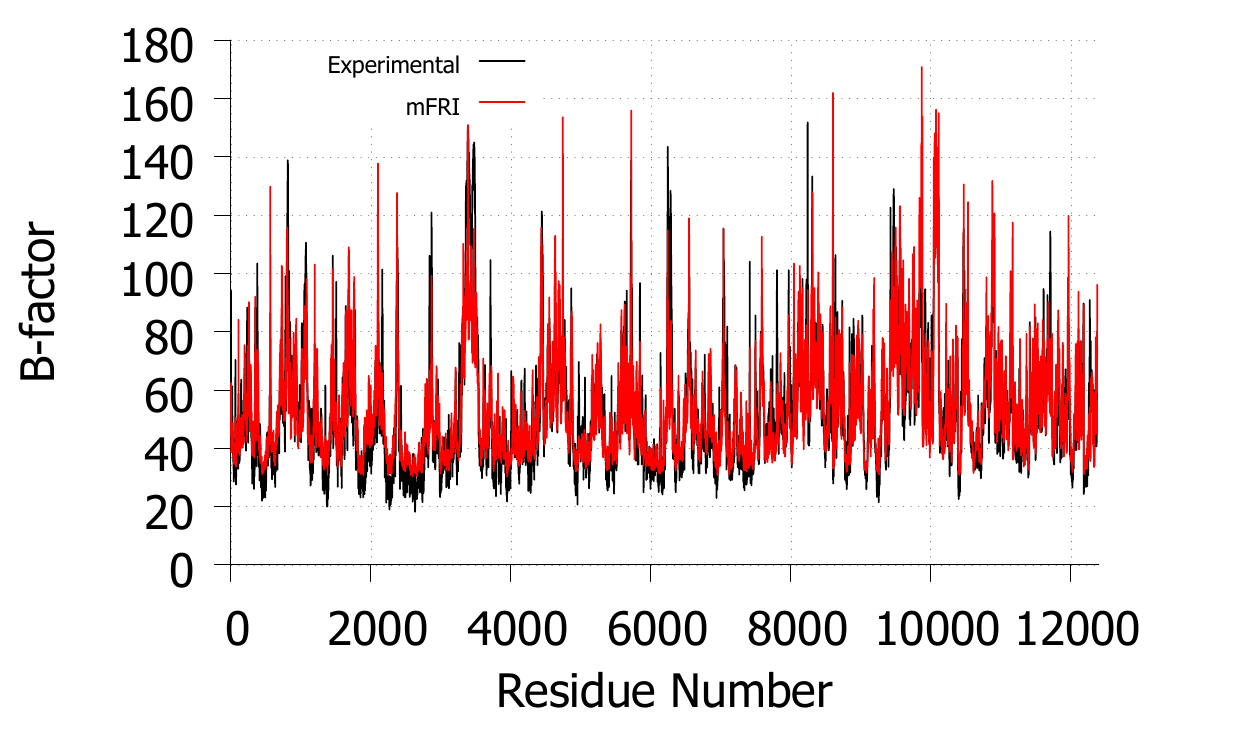}
\label{fig:subfigure2}}
\subfigure[Ribosome 50S subunit PDB ID: 1YIJ]{%
\includegraphics[width=0.40\textwidth]{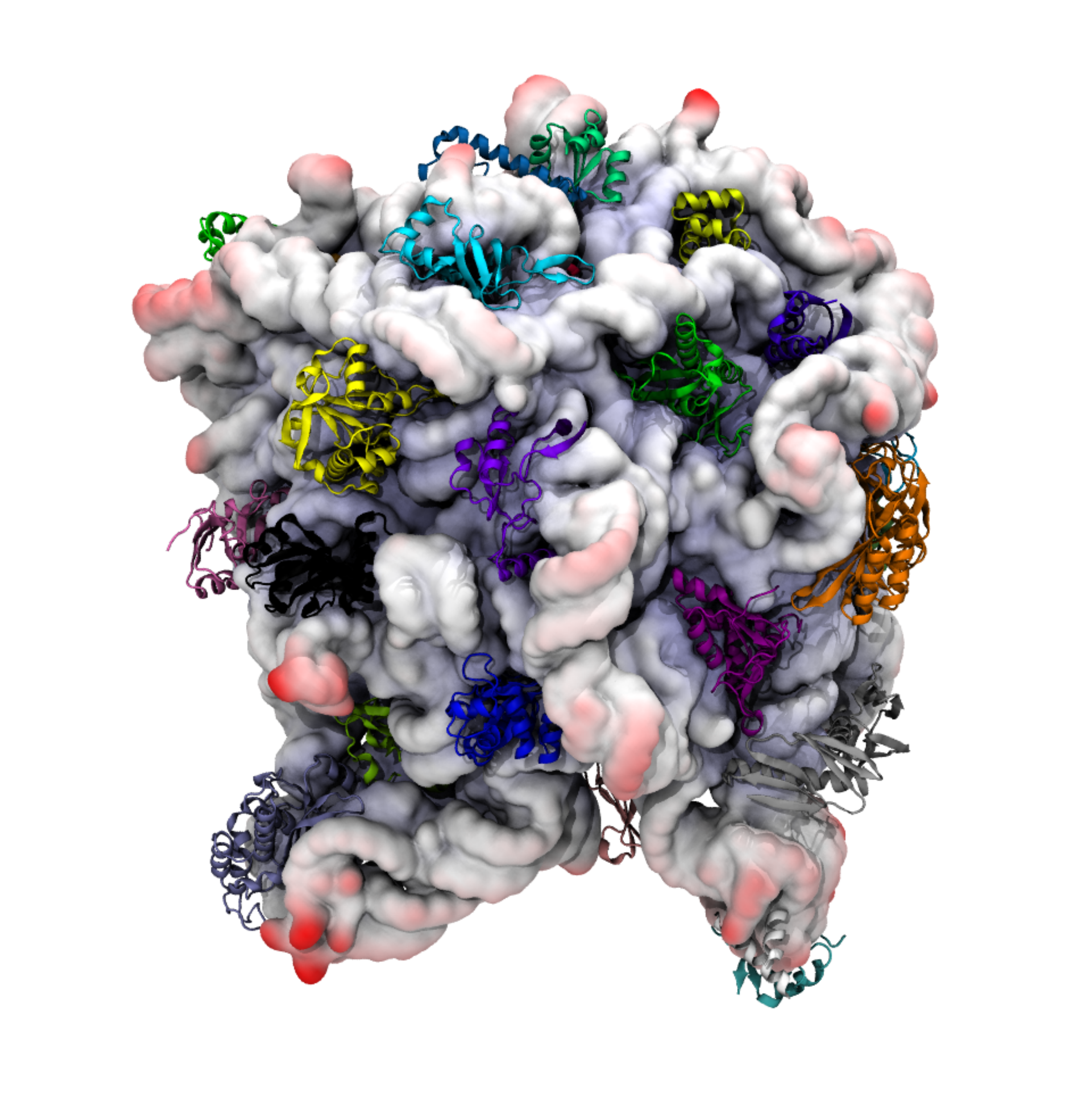}
\label{fig:subfigure3}}
\quad
\caption{Complete ribosome with bound tRNAs (yellow (A site) and green (P site)) and mRNA Shine-Delgarno sequence (orange) PDB ID: 4V4J.  The same correlation coefficients and fitting parameters from mFRI model of protein 1YIJ are used.
A comparison of predicted and experimental B-factor data for  Ribosome 50S subunit PDB ID: 1YIJ. The CC value is 0.85 using the parameter free three-kernel mFRI model.
Nucleic acids are shown as a smooth surface colored by FRI flexibility values (red for more flexible regions) while bound protein subunits are colored randomly and shown in a secondary structure representation.  We achieve a CC value up to 0.85 using parameter free three-kernel mFRI model, i.e., one exponential kernel ($\kappa=1$ and $\eta=15$ \AA) and two Lorentz kernels ($\upsilon=3$, $\eta=3$ \AA ~ and $\upsilon=3$, $\eta=7$\AA).}
\label{ribosome3}
\end{center}
\end{figure}

Some of the largest and most biologically important structures that contain both protein and nucleic acids are ribosomes. Ribosomes are  the protein synthesizers of the cell and connect amino acid into polymer chains. In ribosomes, proteins and RNA interact through intermolecular effects, such as electrostatic interactions, hydrogen bonding, hydrophobic interactions, base stacking and base pairing.  RNA tertiary structures can significantly influence protein-RNA interactions. Ribosomes are primarily composed of RNA with many smaller associated proteins as shown in Fig. \ref{ribosome3}. The top of  Fig. \ref{ribosome3} shows the 50S subunit of the ribosome (PDB ID: 1YIJ) with the nucleic acids in a smooth surface representation with the protein subunits bound and shown in a secondary structure representation.    The set of 64 structures used in our tests contains a number of ribosomal subunits. Due to their multiscale nature, these structures also happen to be among those that benefit the most from using multikernel FRI over single kernel FRI or GNM. For example, in the case of ribosome 50S subunit structure (PDB ID:1YIJ), B-factor prediction with three-kernel FRI yields a CC value of 0.85, while that of single kernel FRI is only around 0.3.  GNM does not provide a good B-factor prediction for this structure either.   The three-kernel mFRI model we used is one exponential kernel ($\kappa=1$ and $\eta=15$ \AA) and two Lorentz kernels ($\upsilon=3$, $\eta=3$ \AA ~ and $\upsilon=3$, $\eta=7$\AA). The comparison between mFRI-predicted and experimental B-factors for  ribosome 50S subunit structure is demonstrated in Fig. \ref{ribosome3}.

By using the fitting coefficients from the above 50S subunit (1YIJ) flexibility analysis, we have obtained flexibility predictions for the entire ribosome (PDB ID:4V4J) as well as many protein subunits and other RNAs that associate with it, see  Fig. \ref{ribosome3}. To avoid confusion, the B-factors for 4V4J are uniquely determined by using not only the same three-kernel mFRI model from the case 1YIJ, and also its fitting parameters, i.e., $a^1=, a^2=, a^3,$ and $b$. Again, the FRI values are mapped by color to the smooth surface of the nucleic acids, however, in these bottom figures the protein subunits are omitted to draw attention instead to the various types of RNA involved in this structure.

\subsection{Anisotropic FRI for conformational motion prediction of an RNA polymerase}\label{sec:RNAP}

RNA polymerase is one of the essential enzymes for all life on Earth as we know it today and possibly from the very beginning of life \cite{Iyer2003,Burton2014old}. Despite this importance, the mechanisms for many of the polymerase's functions are still not well understood on the atomic level. Considerable effort has been spent both experimentally and computationally to understand RNAP polymerase function in more detail but many questions remain. The study of RNA polymerase experimentally or computationally is difficult and often expensive due to the size of the system and variety of molecules involved. The minimal required elements for a bacterial or eukaryotic RNA polymerase include multiple protein subunits, a double stranded DNA molecule, a single stranded RNA molecule, free nucleotides, various ions (Mg$^{2+}$, Zn$^{2+}$, Na$^+$ etc.) and solvent. A typical setup for this system in all-atom molecular dynamics includes ~300,000 atoms when solvated. With this number of atoms and current computer power, it is often not feasible to simulate these molecules on biologically relevant timescales using MD. Perhaps the most popular tool for studying long time dynamics of biomolecules is normal mode analysis (NMA) and its related methods such as the anisotropic network model (ANM). These methods have been successfully used to study protein dynamics for many proteins, however, at their maximum accuracy, their computational complexity is of ${\cal O}(N^3)$, where $N$ is the number of atoms. This is a problem because many cellular functions involve a large number of macromolecules with many thousands to millions of residues to consider. Therefore, future computational studies of biomolecules beyond the protein scale will require methods with better scaling properties such as FRI and aFRI.

\begin{figure}
\begin{center}
\subfigure[RNA Polymerase with closed trigger loop]{%
\includegraphics[width=0.50\textwidth]{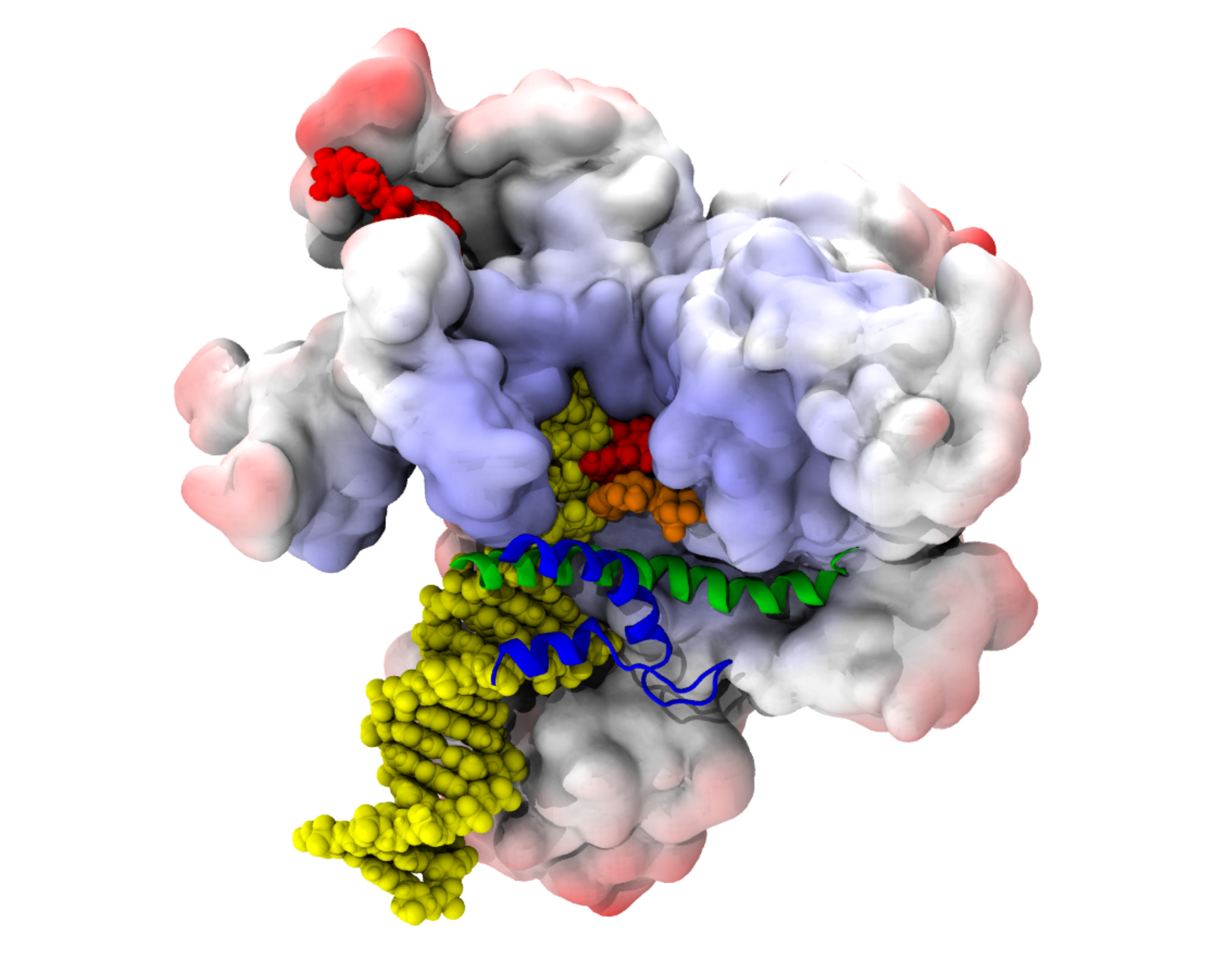}
\label{fig:subfigure4}}
\subfigure[Correlated motion near active site]{%
\includegraphics[width=0.40\textwidth]{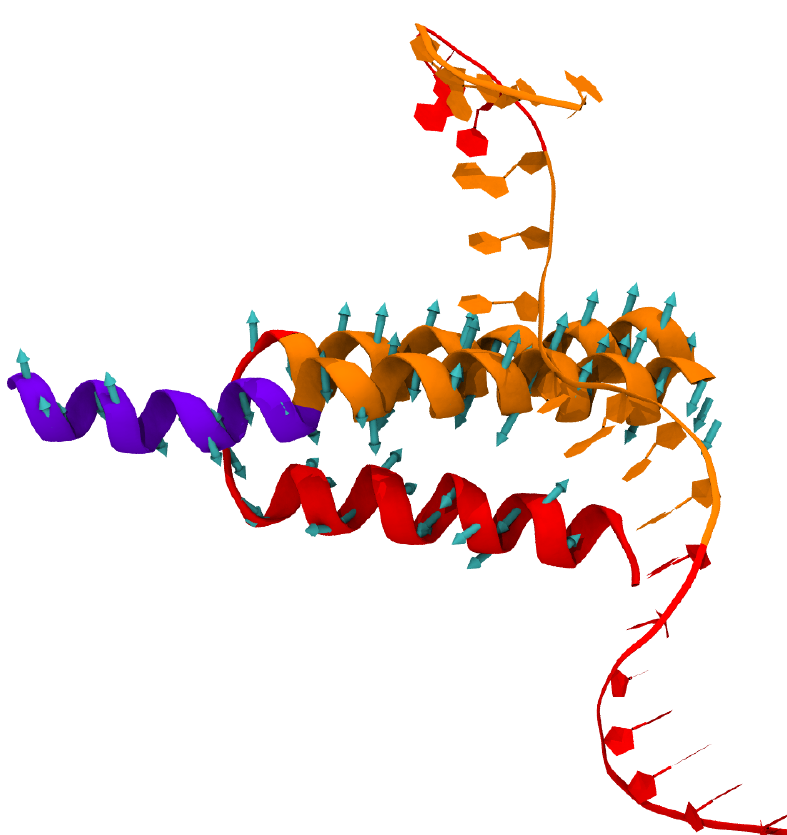}
\label{fig:subfigure5}}
\subfigure[aFRI mode 1 - Open TL]{%
\includegraphics[width=0.40\textwidth]{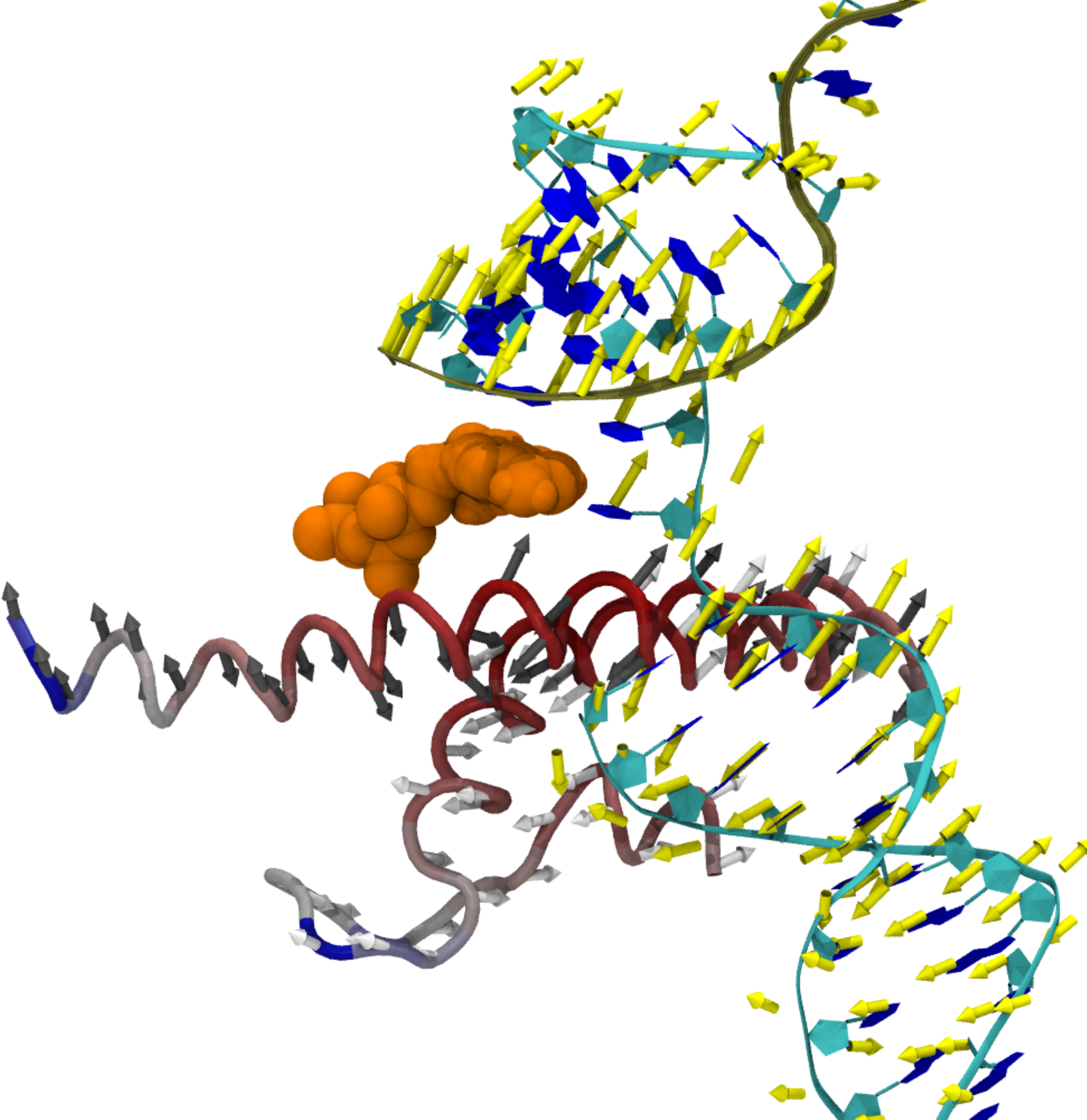}
\label{fig:subfigure6}}
\quad
\subfigure[aFRI mode 1 - Closed TL]{%
\includegraphics[width=0.40\textwidth]{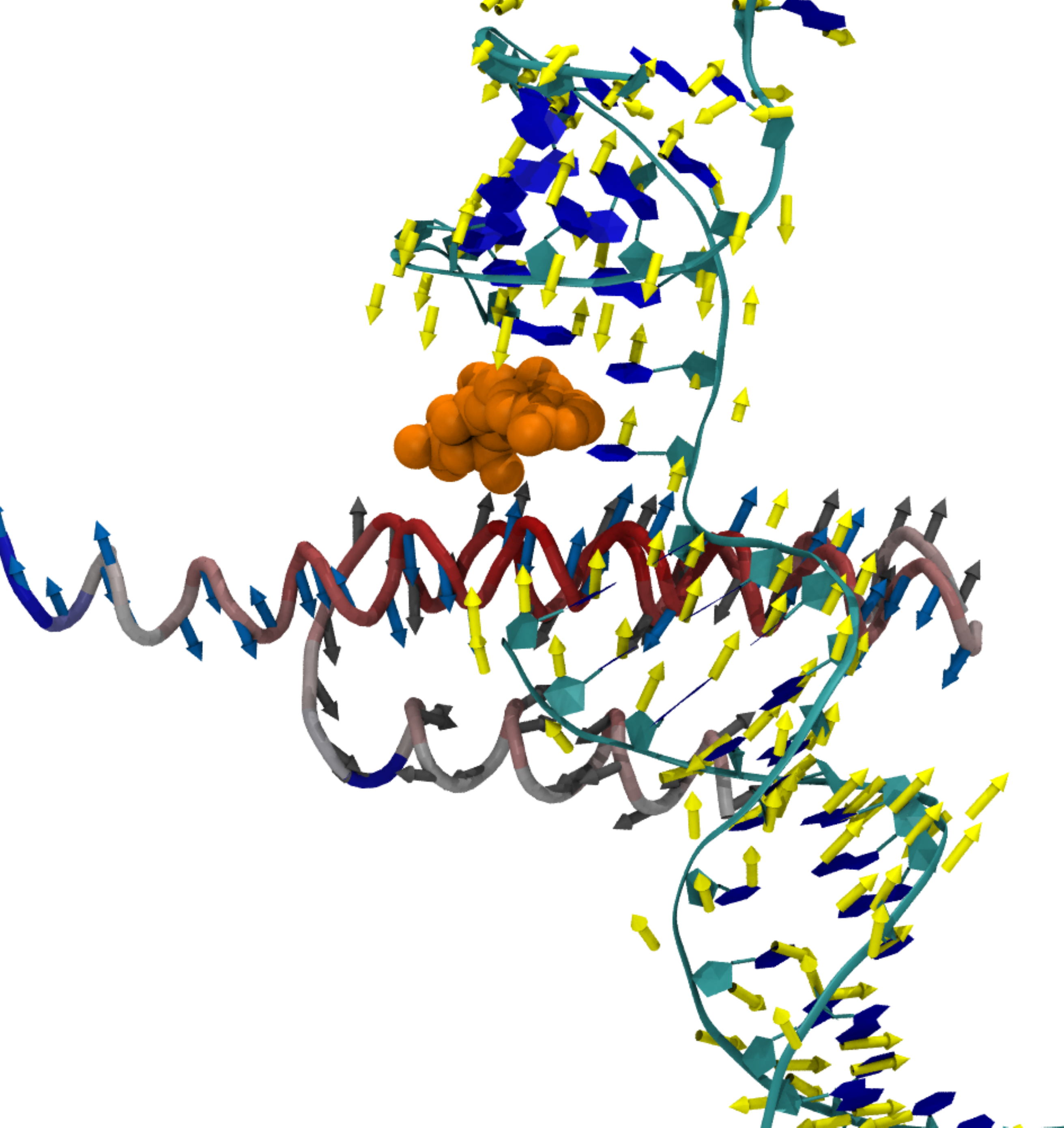}
\label{fig:subfigure7}}
\end{center}
\caption{ The first RNAP local FRI mode  for the bridge helix, trigger loop and nucleic acids from  both open (PDB ID: 2PPB) and close (PDB ID: 2O5J) configurations. Arrows represent the direction and relative magnitude of atomic fluctuations. Arrows for the bridge helix, trigger loop and nucleic acids are pictured as blue, white and yellow, respectively. 
}
\label{RNAP_flex}
\end{figure}

In this example, we use completely local anisotropic FRI to examine correlated motions in regions near the active site of bacterial RNA polymerase, including the bridge helix, trigger loop and nucleic acid chains. We examine the relationship between these components' motions and their contributions to critical functions such as catalysis and translocation.  We use the anisotropic rigidity form in Sec. \ref{sec:arFRI} with the Lorentz kernel ($\upsilon=2$ and $\eta=3$ \AA).  Figure \ref{RNAP_flex}a is a simplified representation of RNA polymerase (PDB ID 2PPB) that shows these important features which are buried in the core of the largest protein subunits, $\beta$ and $\beta'$. The bridge helix and trigger loop, shown in green and blue respectively, are parts of the protein that have been implicated in most of the essential functions of the polymerase. Mutational studies of these regions result in modulation of the polymerase speed and accuracy, both positively and negatively, indicating the regions are important for normal functioning of the enzyme. How these regions aid these functions and how they interact remains an open question. With this demonstration of local aFRI analysis we hope to shed some light on how these essential parts of RNA polymerase work together.

Local aFRI, as described in earlier work, is much less computationally costly than global aFRI or NMA and has been shown to have qualitatively similar results for small to large size single proteins. To further validate the local aFRI method we compare the conclusions from a local aFRI study of RNAP to those of NMA based studies. The RNA polymerase elongation complex is a relatively large system but it is still tenable for NMA methods. NMA has been applied to both bacterial and eukaryotic RNA polymerase in the past  \cite{Van2004, Feig2010rna} which provides us with a point of comparison for our results.

Local aFRI produces three modes of motion sorted from lowest to highest frequency vibration according to eigenvalue as in NMA. In Figure \ref{RNAP_flex} we present findings from the lowest frequency mode effectively focusing on the most dominant motion of each conformation. Two major conformations of RNA polymerase are considered, those with open and closed trigger loop regions (Figures \ref{RNAP_flex}c and \ref{RNAP_flex}d.) A closed trigger loop is one that is completely folded into two parallel alpha helices while an open trigger loop has a region of disordered loop between two shorter helices and is slightly bent away from the bridge helix . The closing or folding of the trigger loop into the closed conformation is assumed to follow binding of an NTP in the active site and to precede catalysis. After catalysis, it is suspected that the trigger loop opens or unfolds to facilitate translocation and permit new NTPs to enter the active site.

The results of aFRI analysis on the effect of trigger loop closing reveal a distinct change in correlated motions in open and closed trigger loop conformations. These changes involve interactions between the bridge helix, the trigger loop and the nucleic acid regions. In Figure \ref{RNAP_flex}b, regions of high correlation are color coded which reveals that the bridge helix is composed of two highly self correlated portions suggesting the presence of a hinge in the bridge helix. In fact, the central portion of the bridge helix has been observed as a kinked or bent helix in a yeast RNAP structure\cite{Wang2006}. Additionally, it is observed that a portion of the bridge helix and the N-terminal helix of the trigger loop are highly correlated  in the closed trigger loop structure only. This set of two helices is situated directly next to the active site and could provide stability to aid catalysis after trigger loop closing.

Additionally, correlation between nucleic acids and protein shows marked differences from the open trigger loop to closed trigger loop structures. The motions indicated in Figures \ref{RNAP_flex}c and \ref{RNAP_flex}d show that the open trigger loop structure is primed to translocate based on the direction of highly correlated motions of the upstream and downstream nucleic acids. By contrast, the closed trigger loop nucleic acid motions are considerably less correlated and not in the direction of translocation. This is the expected relationship as it matches the results from previous biological and NMA studies of RNA polymerase. \cite{ Feig2010rna}

These differences between a closed trigger loop and open trigger loop structure reveal potentially important structural changes that arise as the RNA polymerase switches between open and closed trigger loop conformations during the transition between translocation and catalysis. Specifically, the results for the closed trigger loop conformation suggest the presence of a stabilized catalytic area  which is made of the N-terminal helix of the trigger loop and the bridge helix. The results for the open trigger loop conformation show no such coordination of the active site helices and instead indicates a less defined hinge and coordinated motion in the direction of translocation. Taken together these results provide a potential explanation for how trigger loop opening and closing is correlated with translocation and catalysis respectively.


\section{Concluding remarks}\label{sec:Conclusion}

Protein-nucleic acid complexes are essential to all living organisms. The function of these complexes depends crucially  on their flexibility, an intrinsic property of a macromolecule. However, for many large protein-nucleic acid complexes, such as ribosomes and RNA  polymerases, the present flexibility analysis approaches can be problematic due to their computational complexity scaling of ${\cal O}(N^3)$ and neglecting multiscale effects.

This work introduces the flexibility-rigidity index (FRI) methods \cite{KLXia:2013d,Opron:2014,Opron:2015a} for the flexibility analysis of protein-nucleic acid structures. We show that a multiscale FRI (mFRI) realized by multiple kernels parametrized at multiple length scales is able to significantly outperform the Gaussian network model (GNM) for the  B-factor prediction of a set of 64 protein-nucleic acid complexes \cite{Yang:2006}. The FRI methods are not only accurate, but also efficient, as their computational complexity scales as ${\cal O}(N)$.  Additionally, anisotropic FRI (aFRI), which has cluster Hessian matrices,  offers collective motion analysis for any given cluster, i.e, subunit or domain  in a biomolecular complex.

We apply FRI methods to a large ribosomal subunit (1YIJ) with multiple subunits. We note that both original single-scale FRI and GNM do not work well for this structure.  It is found that  that the multiscale strategy is crucial for the flexibility analysis of multi-subunit structures. The correlation coefficients between FRI predictions and experimental B-factors for 1YIJ improve from 0.3 for single-scale FRI to 0.85 for multiscale FRI. We further use the fitting coefficients obtained from 1YIJ to predict the flexibility of a entire ribosome,  4V4J.  We found that mFRI has an advantage for analyzing large biomolecular complexes due to both higher speeds and accuracy.

We have also demonstrated the utility of the anisotropic FRI (aFRI) for analyzing the translocation of an RNA polymerase, which involves protein, DNA, RNA,  nucleotide substrates and various ions. Both experimental and computational studies of  RNA polymerases are difficult and expensive due to the size and complexity of the biomolecular complex. The molecular mechanism of RNA polymerase translocation is an interesting, open research topic. The present work makes use of localized aFRI to elucidate the synergistic local motions of a bacterial RNA polymerase. Our findings are consistent with those from much more expensive molecular dynamics simulations and normal mode analysis \cite{Feig2010rna,Feig2010rna2}.

{The study of hinges has been an important topic and  much research has been done in the past \cite{hingeprot, flexoracle, hingeatlas, stonehinge, flexprot}. Identification of hinge residues is useful for inferring motion and function when molecules are too large for MD simulation on relevant timescales. Other methods, such as GNM and NMA have been utilized. FRI-based methods could place a significant role in hinge analysis. This aspect will be carefully analyzed in our future work.
}

\vspace{1cm}
\section*{Acknowledgments}

This work was supported in part by NSF grants, IIS-1302285,  and DMS-1160352, NIH Grant R01GM-090208 and Michigan State University Center for Mathematical Molecular Biosciences Initiative. The authors acknowledge the Mathematical Biosciences Institute for hosting valuable workshops.

\vspace{1cm}

\end{document}